\documentclass[aps,pra,twocolumn,superscriptaddress,showkeys]{revtex4-1}
\usepackage{amsmath,amsfonts,amssymb}
\usepackage{graphicx,float,calc}
\usepackage{color,bm}
\usepackage{ulem}
\usepackage{braket}
\usepackage[colorlinks,urlcolor=blue,citecolor=blue,linkcolor=blue]{hyperref}

\begin{document}

\title{Reentrant topological phases and entanglement scalings in moir\'e-modulated extended Su-Schrieffer-Heeger Model}

\author{Guo-Qing Zhang}
\email{zhangptnoone@zjhu.edu.cn}
\affiliation{Research Center for Quantum Physics, Huzhou University, Huzhou 313000, P. R. China}

\author{L. F. Quezada}
\email{lf\_quezada@outlook.com}
\affiliation{Research Center for Quantum Physics, Huzhou University, Huzhou 313000, P. R. China}
\affiliation{Laboratorio de Ciencias de la Informaci\'{o}n Cu\'{a}ntica, Centro de Investigaci\'{o}n en Computaci\'{o}n, Instituto Polit\'{e}cnico Nacional, UPALM, 07700, Ciudad de M\'exico, M\'exico}

\author{Shi-Hai Dong}
\email{dongsh2@yahoo.com}
\affiliation{Research Center for Quantum Physics, Huzhou University, Huzhou 313000, P. R. China}
\affiliation{Laboratorio de Ciencias de la Informaci\'{o}n Cu\'{a}ntica, Centro de Investigaci\'{o}n en Computaci\'{o}n, Instituto Polit\'{e}cnico Nacional, UPALM, 07700, Ciudad de M\'exico, M\'exico}

\date{\today}

\begin{abstract}
Recent studies of moir\'e physics have unveiled a wealth of opportunities for significantly advancing the field of quantum phase transitions. However, properties of reentrant phase transitions driven by moir\'e strength are poorly understood. Here, we investigate the reentrant sequence of phase transitions and the invariant of universality class in moir\'e-modulated extended Su-Schrieffer-Heeger (SSH) model. For the simplified case with intercell hopping $w=0$, we analytically derive renormalization relations of Hamiltonian parameters to explain the reentrant phenomenon. For the general case, numerical phase boundaries are calculated in the thermodynamic limit. The bulk boundary correspondence between zero-energy edge modes and entanglement spectrum is revealed from the degeneracy of both quantities. We also address the correspondence between the central charge obtained from entanglement entropy and the change in winding number during the phase transition. Our results shed light on the understanding of universal characteristics and bulk-boundary correspondence for moir\'e induced reentrant phase transitions in 1D condensed-matter systems.
\end{abstract}

%\keywords{quantum phase transition, moir\'e physics, entanglement entropy, critical behavior, central charge}

\maketitle

\section{Introduction}
Moir\'e physics garnered significant attention in recent years, unveiling exotic properties in both 1D~\cite{gonccalves2024incommensurability,ren2024realizing,PhysRevB.109.045105,zhang2025reentrant} and 2D systems~\cite{cao2018correlated,balents2020superconductivity,Hu2023,Crepel2024,zhang2020twist,Cui2024,dean2013hofstadter,PhysRevLett.122.016401,doi:10.1126/science.aay5533}. The moir\'e pattern, arising from the superposition of two misaligned periodic lattices, emerges a new superlattice with an enlarged periodicity. The band width of a moir\'e system can become nearly flat , hosting compact localized states that promote localization and enable the stabilization of strongly correlated superconductivity and topological insulators~\cite{cao2018unconventional,PhysRevLett.131.016001,nuckolls2020strongly,PhysRevB.107.165114,PhysRevLett.132.036501}. In particular, the uniquely emergent fractional quantum anomalous Hall effect~\cite{PhysRevLett.106.236804,PhysRevX.13.031037,Cai2023,PhysRevB.110.115146,Lu2024,Zhao2025}—that is, the realization of fractional Hall conductance in the absence of an external magnetic field—underscores the significance of moir\'e physics in these systems. Thus, moir\'e systems provides an ideal platform for engineering band structures and novel quantum states. The moir\'e pattern in 2D twisted graphene can be effectively reduced to 1D carbon nanotube lattice, exhibiting significantly modified band structure and distinct characteristics by the emergent moir\'e potential~\cite{PhysRevB.91.035405,PhysRevLett.124.106101}. Other analogues of moir\'e patterns in 1D can be realized through finite incommensurate or commensurate lattice potentials~\cite{PhysRevLett.126.036803}, or finite coupled resonators with varying modulation lengths~\cite{PhysRevLett.130.143801}. Previous studies reveal that the incommensurability and commensurability of moir\'e potentials can enable quasi-fractal charge-density waves and periodic-moir\'e spin-density waves in 1D narrow-band systems~\cite{gonccalves2024incommensurability,zhang2025reentrant}.

Topological phases, usually protected by certain symmetries and characterized by non-trivial boundary states and topological invariants, make the non-trivial properties robust against disorders and local perturbations. These properties have potential applications in the near-future fault-tolerant quantum computation~\cite{RevModPhys.80.1083,he2019topological,Li2024scpma}. Thus, investigating topological phases of matter has been an important field in condensed-matter physical~\cite{RevModPhys.82.3045,RevModPhys.83.1057,Culcer_2020,RevModPhys.90.015001}, and artificially engineered systems~\cite{lu2014topological,roushan2014observation,PhysRevLett.113.050402,PhysRevLett.130.036202,coen2024nonlinear}. In the interacting region, topological phases have been enriched by the fractional quantum Hall states~\cite{PhysRevLett.48.1559} and topological Mott insulators~\cite{PhysRevLett.100.156401,PhysRevA.86.053618,PhysRevB.88.045110,PhysRevB.104.L161118}. Simulations of topological matters have achieved remarkable success in recent experiments, such as the realizations of 1D topological insulator Hamiltonian~\cite{PhysRevLett.110.076401,Song2018}, and the observation of chiral edge states in optical lattices~\cite{doi:10.1126/science.aaa8515,doi:10.1126/science.aaa8736}. Topological phases can be induced by incommensurate quasiperiodic disorders~\cite{PhysRevLett.131.176401,PhysRevA.105.063327,Li2024}, transitioning from a trivial phase, analogous to the random disorder induced topological Anderson insulators~\cite{PhysRevLett.102.136806,PhysRevLett.103.196805,Zhang2020,Gu2023}. Recent work indicates the topological phase can also be driven by commensurate misaligned potentials, exhibiting reentrant phenomenon~\cite{zhang2025reentrant}. However, the critical properties and the profound bulk-boundary correspondence near reentrant transition points in these moir\'e systems remain largely unexplored. Additionally, we aim to probe the transformative potential of moir\'e patterns in the hopping term, unlocking a broader landscape of moir\'e topology, and enabling the emergence of more complex phase transition sequences, including those with winding numbers exceeding 1.

In this work, we address the invariant of universality class in the reentrant sequence phase transitions induced by moir\'e modulation of hopping terms, and present the bulk-boundary correspondence between open-boundary condition (OBC) topological properties and periodic-boundary condition (PBC) entanglement entropy. We start from the extended SSH model and consider a moir\'e pattern modulating the intracell hopping $v$. For the simplified case with intercell hopping $w=0$, the phase transition boundary can be solved analytically, and the reentrant of topological phase transition is explained by the renormalization of Hamiltonian parameters by the moir\'e strength. The universality class is shown to remain invariant across the reentrant sequence of phase transitions, as determined by the critical exponent and dynamical critical exponent, derived from finite-size analysis of the winding number and energy gap, respectively. we than reveal the bulk-boundary correspondence between topological properties and entanglement entropy. The degeneracy of zero-energy edge states under OBCs, and the degeneracy of entanglement spectrum for ground states under PBCs are correlated and interpreted by the entanglement Hamiltonian. We also address the correlation between the central charge obtained from PBC entanglement entropy and the change of OBC winding number during phase transition. We then extend our findings to the general regime of nonzero intercell hopping $w\neq0$. With our rigorous numerical simulations, moir\'e pattern is proven to universally induce the reentrant topological phase transition sequences with winding number larger than 1 while hold the invariant universality class.

%The rest of this paper is organized as follows.  In Sec.~\ref{sec2}, we outline the real- and momentum-space Hamiltonian of the moir\'e-modulated extended SSH model, topological winding number, and many-body entanglement entropy. Section \ref{sec3} is devoted to revealing the universality class and bulk-boundary correspondence of the simplified case $w=0$, where analytical phase boundary and renormalization relation are presented. We then extend our investigation to the general case $w\neq0$ in Sec.~\ref{sec4}, and obtain same results. A brief discussion and conclusion are finally given in Sec.~\ref{secf}.

\section{\label{sec2}Model and Theory}
\begin{figure}[t!]
\centering
\includegraphics[width=0.48\textwidth]{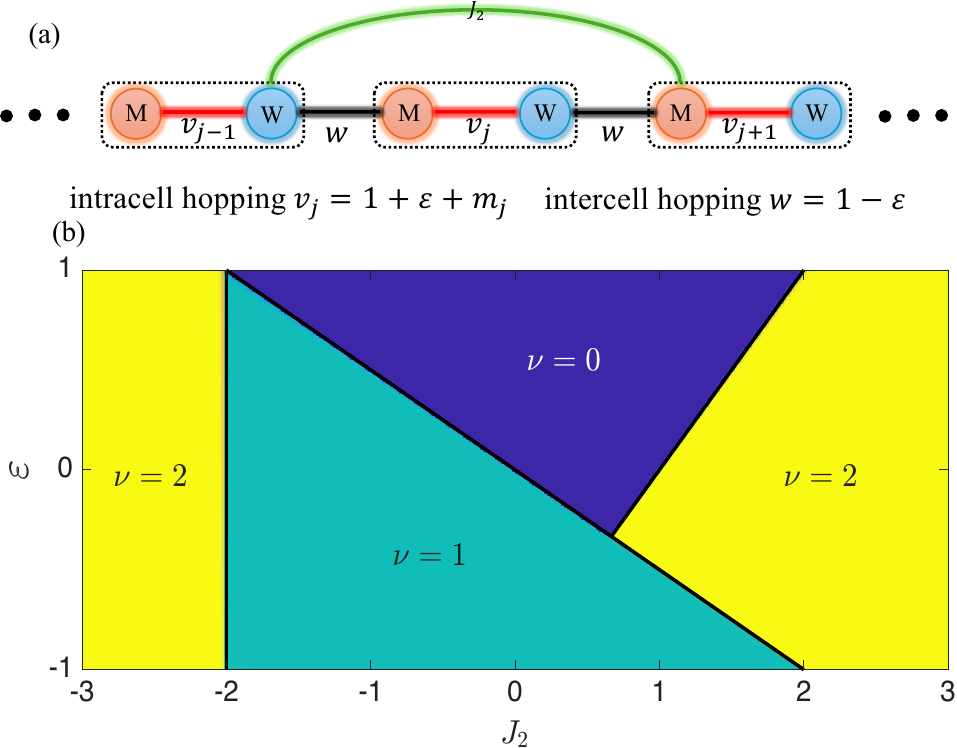}
\caption{(Color online) Illustration and phase diagram of the extended SSH model. (a) Illustration of the investigated extended SSH model with the intracell hopping $v_j$ modulated by moir\'e pattern $m_j=m_o[\cos(2\pi j/a_1)+\cos(2\pi j/a_2)]$. (b) The phase diagram of the extended SSH model with $m_o=0$. The phase diagram consists of three different regions, the topological trivial region with winding number $\nu=0$, the topological non-trivial regions with $\nu=1$ and $\nu=2$.}
\label{figH}
\end{figure}

We consider the extended SSH model with the intracell hopping $v_j$ spatially modulated by the superposition of two periodic hoppings. The original extended SSH model consists of two sublattices M and W in a unit cell with intracell hopping $v=1+\varepsilon$, intercell hopping $w=1-\varepsilon$, and long-range hopping $J_2$. By modulating the intracell hopping $v_j=1+\varepsilon+m_j$, where $m_j=m_o[\cos(2\pi j/a_1)+\cos(2\pi j/a_2)]$ is a moir\'e pattern formed by the superposition of two commensurate periods $a_1$ and $a_2$, the system emerges a new supercell with a periodic of $a_{12}=a_1\times a_2$ unit cells.  The tight-binding Hamiltonian of our considered model reads
\begin{equation}\begin{split}\label{ham}
\hat{H}=&\sum_jv_j(\hat{c}^\dagger_{M,j}\hat{c}_{W,j}+\mathrm{H.c.})+w\sum_j(\hat{c}^\dagger_{W,j}\hat{c}_{M,j+1}+\mathrm{H.c.})\\
&+J_2\sum_j(\hat{c}^\dagger_{M,j+2}\hat{c}_{W,j}+\mathrm{H.c.}),
\end{split}\end{equation}
where $\hat{c}^\dagger_{M(W),j}$ creates a spinless fermion on sublattice M (W) on site $j$. The diagrammatic sketch of our model is illustrated in Fig.~\ref{figH} (a). We consider a system length of $A$ supercells with a total of $L=Aa_1a_2$ unit cells and $N=2L$ sublattice sites. As the chiral symmetry is enough to protect the topology, we refer our model to the AIII class~\cite{RevModPhys.88.035005} with satisfaction $\hat{C}^{-1} \hat{H} \hat{C}=-\hat{H}$, where the chiral operator $\hat{C}=\mathbf{I}_L\otimes\hat{\sigma_z}$, $\mathbf{I}_L$ is an $L$-rank identity matrix, and $\hat{\sigma_z}$ is the Pauli matrix. For PBCs, the Hamiltonian (\ref{ham}) can be Fourier transformed to momentum space. The moir\'e pattern term leads to
\begin{equation}
\sum_j\cos \left(\frac{2\pi}{a_\alpha}\right)\hat{c}^\dagger_{M,j}\hat{c}_{W,j}=\frac{1}{2}\sum_k\left(\hat{c}^\dagger_{M,k}\hat{c}_{W,k+\frac{2\pi}{a_\alpha}}+\hat{c}^\dagger_{W,k}\hat{c}_{M,k+\frac{2\pi}{a_\alpha}}\right),
\end{equation}
where $\alpha=1,2$. By expanding the momentum-space Block basis $\ket{BK}=[\hat{c}_{M,k+0\pi/{a}_{12}}\  \hat{c}_{M,k+2\pi/{a}_{12}} \ \cdots\  \hat{c}_{W,k+0\pi/{a}_{12}}\  \hat{c}_{W,k+2\pi/{a}_{12}}\  \cdots]^\mathrm{T}$ in the reduced moir\'e Brillouin zone (MBZ) $k\in [0,2\pi/a_{12})$~\cite{zhang2025reentrant}, we can write the Bloch Hamiltonian $\hat{H}(k)$ as
\begin{equation}\begin{split}\label{hamk}
\hat{H}(k)=&[\hat{\sigma}_x\otimes(v+\mathbf{X})+\hat{\sigma}_y\otimes \mathbf{Y}]\\
&+\hat{\sigma}_x\otimes\frac{m_o}{2}(\mathbf{E}+\mathbf{F}),
\end{split}\end{equation}
where $\mathbf{X},\mathbf{Y},\mathbf{E},\mathbf{F}$ are $a_{12}\times a_{12}$ square matrices, with $\mathbf{X}=\mathrm{diag}[\cdots,w\cos(k+k_l)+J_2\cos(2k+2k_l),\cdots]$, $\mathbf{Y}=\mathrm{diag}[\cdots,w\sin(k+k_l)+J_2\sin(2k+2k_l),\cdots]$, $\mathbf{E}(i,j)=\delta_{j,(i+a_2)\%a_{12}}+\delta_{j,(i+a_{12}-a_2)\%a_{12}}$, $\mathbf{F}(i,j)=\delta_{j,(i+a_1)\%a_{12}}+\delta_{j,(i+a_{12}-a_1)\%a_{12}}$. The operator $\%$ stands for mod, and $k_l=2\pi (l-1)/a_{12}$ for $l=1,2,\cdots,a_{12}$. It's obvious that the Bloch Hamiltonian is in off-diagonal block form, and the energy spectrum is separated to $2a_{12}$ subbands.

Our model obeys $\mathbb{Z}$ topological index and can be characterized by the integer winding number $\nu$. Under the OBC, the translational symmetry is broken and we can use the real-space winding number to characterize the bulk topology~\cite{PhysRevLett.113.046802}
\begin{equation}
\nu=\frac{1}{L'}\mathrm{Tr}'(\hat{C}\hat{P}[\hat{P},\hat{X}]),
\end{equation}
where the projector $\hat{P}=\sum_{l=1}^L(\ket{\psi_l}\bra{\psi_l}-\hat{C}\ket{\psi_l}\bra{\psi_l}\hat{C}^{-1})$ sums the lowest half wave functions $\ket{\psi_l}$ of Hamiltonian (\ref{ham}) for our half-filling case, $\hat{X}=\mathrm{diag}(1,1,2,2,\cdots,L,L)$ is the coordinate operator, and $\mathrm{Tr}'$ stands for the trace per volume over the center internal $L'=L/2$ matrix elements to avoid boundary effects. On the other hand, the momentum-space winding number under the PBC $\nu_k$ can be derived from the off-diagonal Bloch Hamiltonian (\ref{hamk}) where the off-diagonal block $q(k)=v+w\mathbf{X}+m_o(\mathbf{E}+\mathbf{F})/2-iw\mathbf{Y}$ is an ${a}_{12} \times {a}_{12}$ matrix. The momentum-space winding number is then obtained by integral over the reduced MBZ~\cite{zhang2025reentrant}
\begin{equation}\label{nu_k}
\nu_k=\frac{1}{2\pi i}\int_0^{\frac{2\pi}{{a}_{12}}} \mathrm{d}k~ \mathrm{Tr}\left[q(k)^{-1}\partial_kq(k)\right].
\end{equation}
In general, an analytical solution of $\nu_k$ for arbitrary $a_1$ and $a_2$ is not possible and numerical integration is usually needed to obtain $\nu_k$.

To calculate the many-body ground state entanglement entropy, we adopt the single-particle
correlation matrix method from Refs.~\cite{Peschel_2009,PhysRevResearch.4.043164}, and briefly recap key ingredients in the following. At the half-filling case, the many-body ground state $\ket{\Psi}$ can be constructed from the low laying half single-particle wavefunctions, and the density matrix is $\hat{\rho}=\ket{\Psi}\bra{\Psi}$. We divide the system into two parts, A and B. Part A consists of a continuous series of $l$ unit cells starting from the first unit cell in the system, while part B contains the remaining $L-l$ unit cells. For an equal bipartition, part A includes the left $l=L/2$ unit cells and part B is the right $L/2$ unit cells. Here, $l$ denotes the bipartition length. By tracing out the degrees of freedom in subsystem B, with the remaining reduced density matrix $\hat{\rho}_A=\mathrm{Tr}_B\hat{\rho}=e^{-\hat{H}_A}/\mathrm{Tr}e^{-\hat{H}_A}$. The entanglement Hamiltonian $\hat{H}_A$ admits a spectral decomposition $\hat{H}_A=\sum_l \xi_l \hat{\phi}_l^\dagger\hat{\phi}_l$. Directly computing the entanglement entropy and entanglement spectrum from $\hat{H}_A$ is challenging. Instead, the single-particle correlation matrix $C_{mn}=\braket{\Psi|\hat{c}_m^\dagger\hat{c}_n|\Psi}$ is introduced~\cite{PhysRevResearch.4.043164}. The correlation matrix can be cast into the form of single-particle wavefunctions $C_{mn}=\sum_{l=1}^{L}\braket{n|\psi_l}\braket{\psi_l|m}$. The one-to-one correspondence between entanglement Hamiltonian $\hat{H}_A$ spectrum $\{\xi_l\}$ and correlation matrix $C_{mn}$ spectrum $\{\zeta_l\}$ reads
\begin{equation}\label{es}
\zeta_l=\frac{1}{e^{\xi_l}+1},\ \ \ \ l\in A,
\end{equation}
and the entanglement entropy $\mathcal{S}=-\mathrm{Tr}(\hat{\rho}_A\ln\hat{\rho}_A)$ is given by
\begin{equation}\label{ee}
\mathcal{S}=-\sum_l\zeta_l\ln \zeta_l+(1-\zeta_l)\ln(1-\zeta_l),\ \ \ \ l\in A.
\end{equation}
We denote the eigenspectrum $\{\zeta_l\}$ as the entanglement spectrum in later discussion.

When the moir\'e strength $m_o=0$, the Bloch Hamiltonian reduced to a two-band system, and the off-diagonal element is a scalar $q(k)=v+we^{-ik}+J_2e^{-2ik}$. The phase diagram of this case under OBC is presented in Fig.~\ref{figH} (b). The PBC phase boundaries [black solid lines in Fig.~\ref{figH} (b)] can analytically solved from $q(k)=0$, which leads to three solutions: $J_2-2=0$ for $k=0$, $2\varepsilon+J2=0$ for $k=\pi$, and $\varepsilon+1-J_2=0$ for $k=\arccos[(\varepsilon-1)/2J_2]$. The phase diagram consists of three different regions, two topological phases with different winding numbers $\nu=1,2$, and one trivial region $\nu=0$. For clarity of the following investigation while preserving the generality of our findings, we only consider the commensurate period pair $a_1=3$ and $a_2=7$.  Consistent with our previous work~\cite{zhang2025reentrant}, the reentrant phenomenon, invariant of universality class, and bulk-boundary correspondence hold for many other pairs of $a_1$, $a_2$.

\section{Intercell hopping $w=0$\label{sec3}}
For the simplified Hamiltonian where $w=0$ ($\varepsilon=1$), we can analytically derive the phase boundary and renormalization relation. The topological nature of our chiral-symmetric model is revealed by real-space winding number under OBC and entanglement entropy under PBC. We also reveal the bulk-boundary correspondence between zero-energy modes under OBC and entanglement spectrum under PBC.

The analytical phase boundary of this moir\'e-modulated SSH model with $w=0$ can be obtained from the gap-closing nature at topological phase transition point, where the localization length $\Lambda$ of zero-energy modes will diverge under OBC and thus $\Lambda^{-1}=0$~\cite{PhysRevLett.113.046802}. To solve this, we first write the OBC Hamiltonian in matrix form
\begin{equation}\label{H0}
	\hat{H}=\left(\begin{array}{ccccccccc}
		0 & v_1 & 0 & 0  & 0     & 0 &  0     &    0   &   \\
		v_1 & 0 & 0 & 0  & J_2 &  0&      0 &     0  & \cdots  \\
		0 & 0 & 0 & v_2  & 0     & 0 & 0     & 0    &  \\
		0 & 0 & v_2 & 0  & 0     & 0 & J_2 &     0 &  \\
		   &    &    &     &\vdots&    &         &        & \ddots
	\end{array}\right).
\end{equation}
The zero-energy eigenstate is assumed to be $\psi=\{\psi_{A,1},\psi_{B,1} \cdots \psi_{A,L}, \psi_{B,L} \}^T$, and satisfies ${\hat{H}}\ket{\psi} =0$. The recursive relations of the zero-energy eigenstate read
\begin{align}
	v_j\psi_{B,j}&=0,\\
	v_j\psi_{A,j}+J_2\psi_{A,j+2}&=0.
\end{align}
It's evident that $\psi_{B,L}=0$, with the solutions for $\psi_{A,L}$ decoupling into two distinct equations for even and odd sites in the thermodynamic limit $L\rightarrow \infty$, respectively:
\begin{align}
	\psi_{A,2L-1}&=\prod_{j=1}^{L-1}\frac{v_{2j-1}}{J_2}\psi_{A,1},\\
	\psi_{A,2L}&=\prod_{j=1}^{L-1}\frac{v_{2j}}{J_2}\psi_{A,2}.
\end{align}
The inverse of localization length for odd and even sites can be obtained as
\begin{align}
\Lambda^{-1}&=\frac{1}{L} \left |  \ln \left|\frac{\psi_{A,2L-1}}{\psi_{A,1}} \right| \right |,\\
\Lambda^{-1}&=\frac{1}{L} \left |  \ln \left|\frac{\psi_{A,2L}}{\psi_{A,2}} \right| \right |.
\end{align}
Considering the periodicity of $v_j$, these two equations have the same solution for $\Lambda^{-1}=0$ in the thermodynamic limit
\begin{equation}\label{loc}
\ln \prod_{j=1}^{a_{12}}\left|2+m_o\left[\cos\left(\frac{2\pi j}{a_1}\right)+\cos\left(\frac{2\pi j}{a_2}\right)\right]\right|^{\frac{1}{a_{12}}}=\ln J_2,
\end{equation}
which means two pairs of divergent localized states emerge at the transition point and the topological phase has winding number $\nu=2$.

\begin{figure}[t!]
\centering
\includegraphics[width=0.48\textwidth]{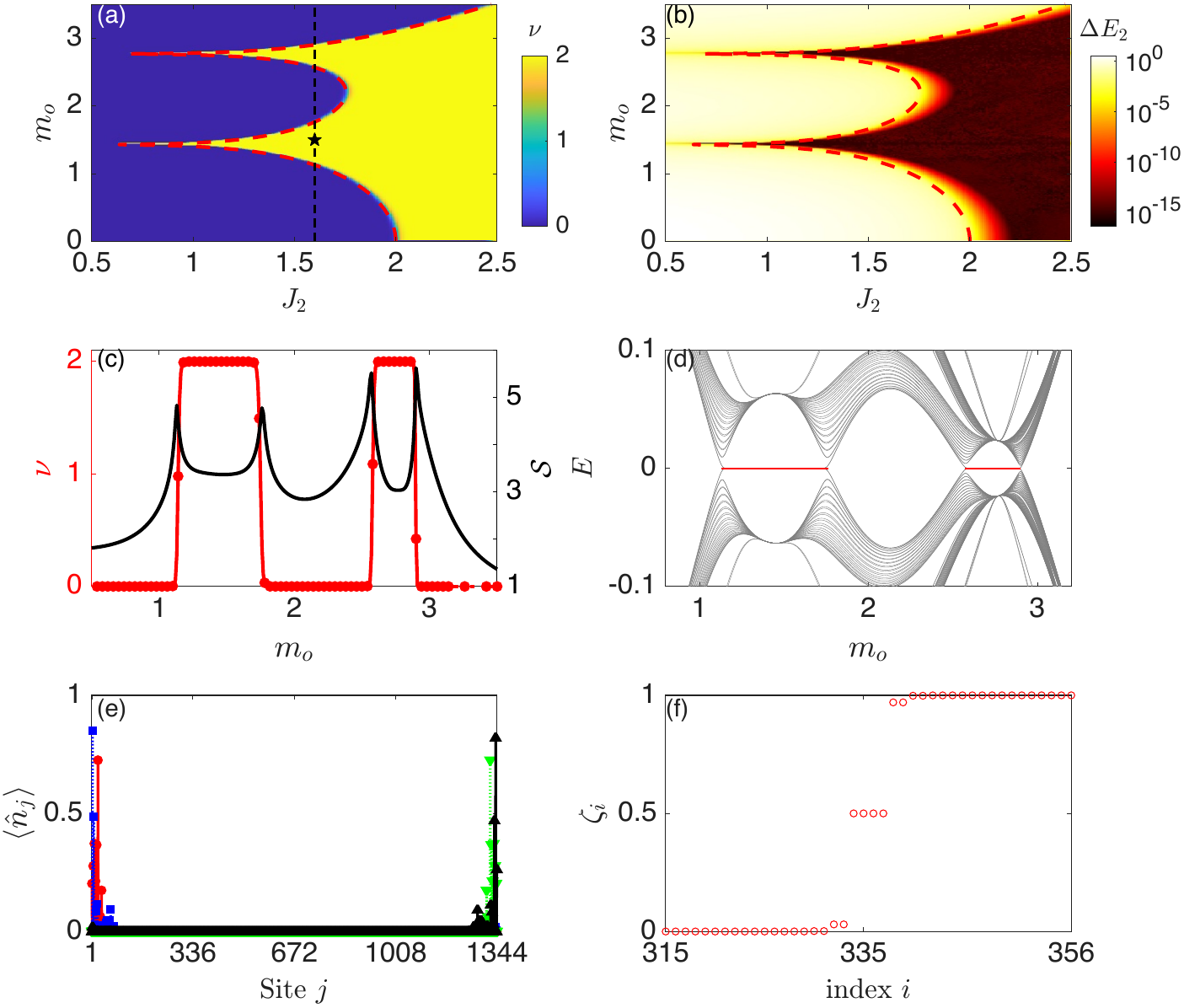}
\caption{(Color online) Characterization of reentrant physics with intercell hopping $w=0$. (a) The topological phase diagram of moir\'e-modulated SSH model. Red dashed curve is the analytical phase boundary obtained from the divergence of localization length. The vertical black dashed line corresponds to the cutting line in (c), and the black pentagram mark denotes the chosen parameters in (e,f). (b) The second energy gap $\Delta E_2$ plotted as functions of $m_o$ and $J_2$. (c) Real-space winding number $\nu$ and entanglement entropy $\mathcal{S}$ plotted as functions of moir\'e strength $m_o$ for $J_2=1.6$. (d) The energy spectrum versus $m_o$ for $J_2=1.6$, with zero-energy modes drawn in red. (e) Density distributions $\braket{\hat{n}_j}$ of four localized edge states for $J_2=1.6$ and $m_o=1.5$. (f) Entanglement spectrum $\zeta_i$ for $J_2=1.6$ and $m_o=1.5$. Other parameters are $\varepsilon=1$, $A=32$, $a_1=3$ and $a_2=7$.}
\label{figpd}
\end{figure}

We plot the phase diagram in Fig.~\ref{figpd} (a) revealed by real-space winding number, along with the analytical phase boundary solved from Eq.~(\ref{loc}), which are consistent with each other. In the $w=0$ limit, the extended SSH model hosts trivial to $\nu=2$ topological phase transition, with two pairs of zero-energy modes emergent at the transition point, and both first and second energy gaps vanish in the $\nu=2$ region topological region. The first energy gap $\Delta E_1$ and the second energy gap $\Delta E_2$ are given as
\begin{align}
\Delta E_1&=E_{L+1}-E_L,\\
\Delta E_2&=E_{L+2}-E_L.
\end{align} Thus, we present the second energy gap $\Delta E_2$ in the $m_o$-$J_2$ parameter space in Fig.~\ref{figpd} (b). As the finite size effect for our numerical simulations under OBC, the second energy gap gradually closes after $\nu=0\rightarrow \nu= 2$ phase transition. For various parameters, the system exhibits reentrant topological phase transition induced by the moir\'e strength $m_o$. We depict the reentrance of OBC topological invariant $\nu$ for $J_2=1.6$ in Fig.~\ref{figpd} (c) as well as the PBC entanglement entropy $\mathcal{S}$. The corresponding energy spectrum is plotted in Fig.~\ref{figpd} (d), with zero energies highlighted in red. In the critical region, the growth of entanglement entropy follows a logarithmic law, scaling as $\mathcal{S}\sim \ln L$ for 1D free fermion systems. At the transition point, the entanglement diverges when $L\rightarrow \infty$. As shown in Fig~\ref{figpd} (c), all topological phase transitions are accompanied by a distinct change in the entanglement value. The correspondence of OBC winding number and PBC entanglement will be discussed latter, here we reveal the correspondence of OBC zero-energy edge modes and PBC entanglement spectrum. The conventional bulk-boundary correspondence involves the OBC edge states with eigenenergy $E=0$. As analytically demonstrated previously, the $\nu=2$ topological region hosts two pairs of localized states, each comprising one left-localized and one right-localized state. We present four zero energy states, whose density distributions are localized at one edge of the system, in Fig.~\ref{figpd} (e). What's more, we reveal that the four degenerate zero energy states under OBC correspond to four degenerate eigenmodes in the entanglement spectrum, which is shown in Fig.~\ref{figpd} (f). From Eq.~(\ref{es}), the eigenvalue of $\xi_i=0$ corresponds to $\zeta_i=1/(e^0+1)=0.5$, and the four degenerate entanglement spectrum values correspond to four zero-energy modes~\cite{PhysRevLett.133.026601}. The funding of $\xi_i=0$ in the topological region is also supported by recent work~\cite{kumar2025entanglemententropyprobetopological}, which reveals that the entanglement entropies for half-filled and near-half-filled ground states are identical in topological phases.

The reentrant topology can be interpreted as the renormalization of Hamiltonian parameters driven by the moir\'e strength. Specifically, the winding number quantifies how many times the momentum-space Hamiltonian encircles the origin~\cite{PhysRevA.97.052115}, and the phase transition point corresponds to $\det q(k)$ crossing the original point in the complex plane when momentum $k$ varies in the reduced MBZ.
For the critical point $m_o=1.1283(3)$, $\det q(k)$ crosses the origin twice, revealing the $\nu=0\rightarrow\nu=2$ phase transition. By solving $\det q(k)=0$, one can find the renormalization relation. However, an analytical solution for $a_{12}=21$ is not feasible within our extended SSH model. We use the localization approach to derive the analytical phase boundary. Our calculation ensures that the vanishing of $\Lambda^{-1}$ corresponds to the change of winding number $\nu_k$, which is exactly the phase transition point. The solution for $\Lambda^{-1}=0$ is given by Eq.~(\ref{loc}), where we can assume the left-hand side as the renormalization of intracell hopping $v$ by the moir\'e strength
\begin{equation}
\tilde{v}=\prod_{j=1}^{a_{12}}\left|v+m_o\left[\cos\left(\frac{2\pi j}{a_1}\right)+\cos\left(\frac{2\pi j}{a_2}\right)\right]\right|^{\frac{1}{a_{12}}}.
\end{equation}
The analytical phase boundary is determined by solving $\ln\tilde{v}=\ln J_2$, which yields $2a_{12}$ complex roots, making the multiple trivial-topological transitions possible in the real root subspace. We incorporate the long-range hopping term $J_2$ to enrich the possible range of topological phases and transition sequences, and introduce the moir\'e pattern to induce the appearance of reentrant phase transitions. For sufficiently large $J_2$ and small $m_0$, the moir\'e-renormalized intracell hopping $\tilde{v}$ is always smaller than $J_2$, thereby placing the system in the winding number $\nu=2$ phase. Consequently, increasing $J_2$ requires a much larger $m_0$ to drive the $\nu=2\rightarrow \nu=0$ phase transition.

\begin{figure}[t!]
\centering
\includegraphics[width=0.48\textwidth]{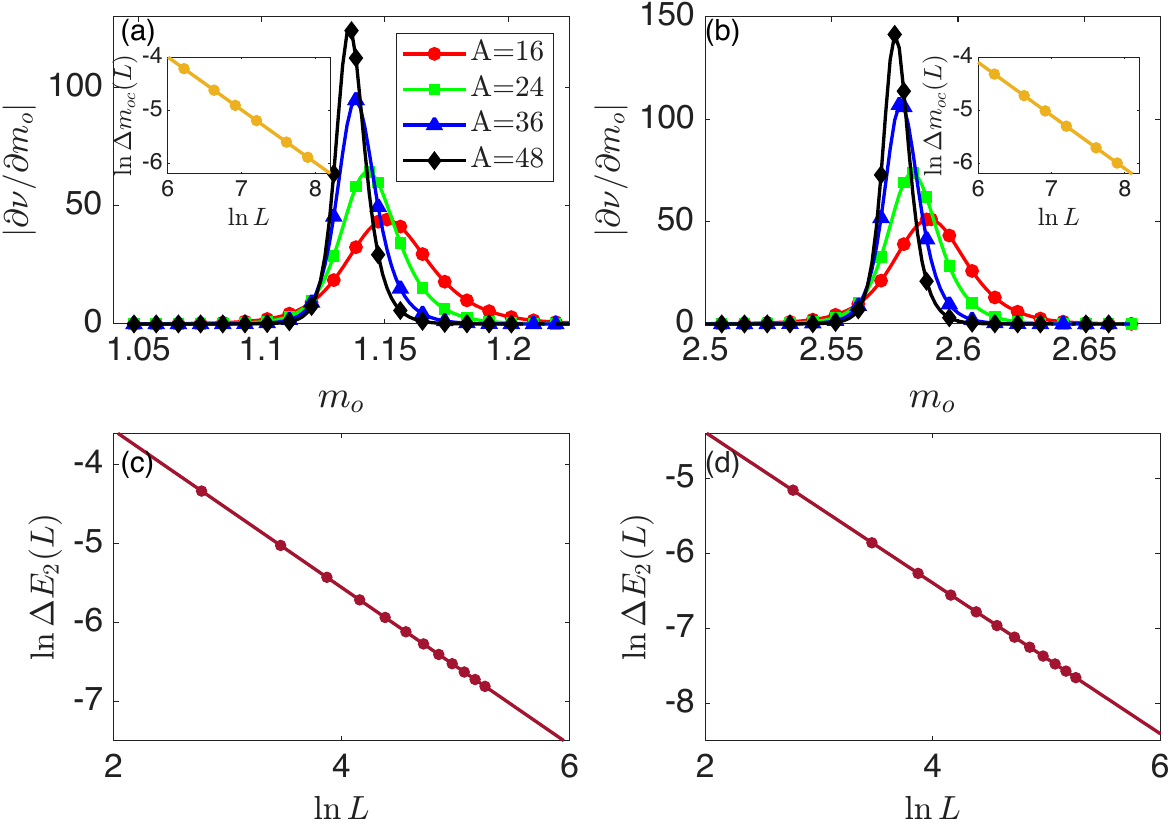}
\caption{(Color online) Critical behaviors with intercell hopping $w=0$. (a) $\partial \nu / \partial m_o$ as a function of $m_o$ for various system sizes near the first trivial-to-topological transition point $m_{oc1}=1.1283(3)$. The finite-size critical point $m_{oc}^{(L)}$ is revealed by the peak position of each curve. Inset panel shows the $\log$-$\log$ plot of system size $L$ and $\Delta m_{oc}(L)=|m_{oc}(L)-m_{oc1}|$. (b) $\partial \nu / \partial m_o$ near the second trivial-to-topological transition point $m_{oc2}=2.5685(7)$. Inset is the $\log$-$\log$ plot of system size $L$ and $\Delta m_{oc}(L)=|m_{oc}(L)-m_{oc2}|$. (c) $\mathrm{Log}$-$\log$ plot of the second energy gap versus system size $L$ near the first (c) and second (d) trivial-to-topological transition points, respectively. Other parameter are $\varepsilon=1$, $a_1=3$, $a_2=7$, and $J_2=1.6$.}
\label{figscal}
\end{figure}

To demonstrate the invariance of universality class for reentrant transitions induced by the moir\'e pattern, we conduct a scale-invariant analysis of the real-space winding number and energy gap across various finite-size systems. Similar analysis of physical quantities such as the fidelity susceptibility and quantum entanglement has been widely used to investigate the critical exponent of quantum phase transitions~\cite{Osterloh2002,PhysRevB.77.245109,PhysRevB.81.064418}. For phase transition with finite-size effect, the critical point is identified by the first derivative of physical quantities, and progressively converges to the true critical point in the thermodynamic limit. Figures~\ref{figscal} (a) and (b) show the numerical results of $\partial\nu/\partial m_o$ as functions of $m_o$ for various system sizes near the first and second trivial-to-topological phase transitions, respectively. As we can see, the finite-size critical point $m_{oc}^{(L)}$, which corresponds to the peck of each curve, approach to the true topological transition point obtained analytically. It is found that the distances from $m_{oc}^{(L)}$ to $m_{oc}$ are well fitted by the power-law scaling form \cite{Chen2016,PhysRevB.95.075116,PhysRevB.110.045119,zhang2025reentrant}
\begin{equation}
	\Delta m_{oc}(L)=|m_{oc}^{(L)}-m_{oc}|\propto L^{-\mu},
\end{equation}
where $\mu$ is the critical exponent related to the universality class.  Distances $\Delta m_{oc}(L)$ as functions of the system size $L$ are $\log$-$\log$ depicted in inset panels of Figs.~\ref{figscal} (a) and (b), which reveal the critical exponent $\mu=0.9984(2)$ for the first transition point and $\mu=1.0022(4)$ for the second transition point. The energy gap closing also indicates a phase transition, and the dynamical critical exponent $z$ can be derived from the scaling relation~\cite{PhysRevB.110.024207}
\begin{equation}
\Delta E_2\propto L^{-z},
\end{equation}
where we focus on the second energy gap $\Delta E_2$ because the first energy gap is always zero for $\nu=1\rightarrow\nu=2$ transitions. Near the first and second trivial-to-topological phase transitions, we present the scaling of energy gap $\Delta E_2$ while varying system size $L$ in Figs.~\ref{figscal} (c) and (d), respectively. Both $\log$-$\log$ plots indicate well fitted dynamical critical exponents $z=0.9977(9)$ and $z=1.0058(6)$ for the first and second phase transitions.
As indicated by the renormalization relation, the moir\'e pattern do no change the universality class of the extended SSH model. We numerically show that these reentrant topological transitions belong to the same universality class with $\mu\approx1$ and $z\approx1$, and are different from the disorder-driven Anderson topological transitions with
$\mu\approx2$~\cite{PhysRevB.109.L201102}.

\begin{figure}[t!]
\centering
\includegraphics[width=0.48\textwidth]{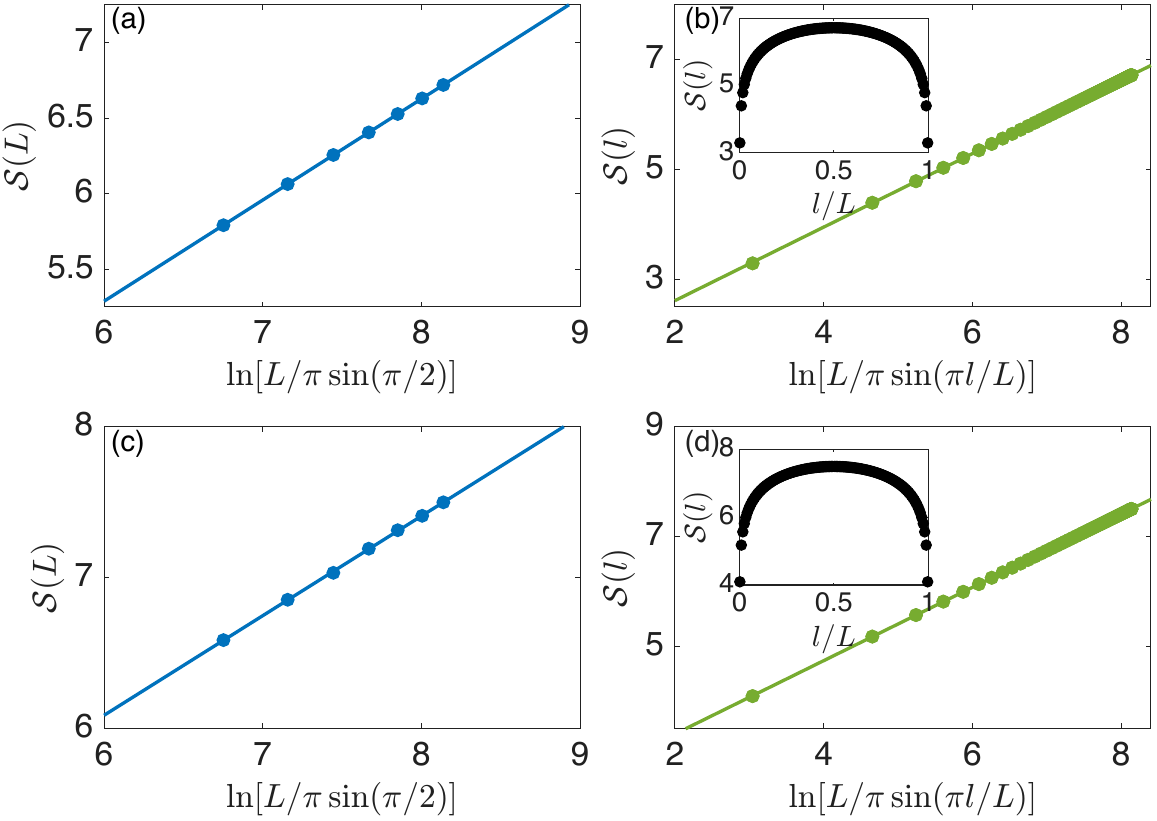}
\caption{(Color online) Entanglement scalings with intercell hopping $w=0$. Entanglement entropy $\mathcal{S}(L)$ scales  as a function of rescaled system size $L$ in the first (a) and second (c) trivial-to-topological phase transition points.  Entanglement entropy $\mathcal{S}(l)$ scales as a function of rescaled bipartition length $l$ or the first (b) and second (d) phase transition points. Inset panels are $\mathcal{S}(l)$ against unscaled $l$. All solid lines are linear fittings which indicate the central charge $c\approx 2$. $A=512$ for (b,d), other parameter are $\varepsilon=1$, $a_1=3$, $a_2=7$, and $J_2=1.6$.}
\label{figee}
\end{figure}

The bulk entanglement entropy spectrum is revealed to encode the information of topologically degenerate zero modes, and we further investigate the relation between bulk entanglement entropy and topological invariant. Based on methods of conformal field theory (CFT), the entropy at a critical point is found to be~\cite{PasqualeCalabrese_2004,Calabrese_2009}
\begin{equation}\label{eescal}
\mathcal{S}=\frac{c}{3}\log\left(\frac{L}{\pi}\sin\frac{\pi l}{L}\right)+c^\prime_1,
\end{equation}
where $c$ is the central charge of the corresponding CFT, $l$ is bipartition length, and $c^\prime_1
$ is a non-universal constant. The central charge $c$, which can be extracted from entanglement scaling, is one of the most important quantities characterizing the CFT. It quantifies the degrees of freedom associated with the edge states in free fields, which we latter connect to the change in winding number during the phase transition. The entanglement entropy given by Eq.~(\ref{ee}) can be computed to any bipartition length $l$ for a certain $L$, and the relation between the central charge $c$ and entanglement entropy $\mathcal{S}$ can be derived by scaling system size $L$ or bipartition length $l$. While varying the system size $L$, we focus on the half-bipartite entanglement entropy with $l=L/2$. The entanglement entropy grows with the increase of $L$ as the volume law implied. For the first and second trivial-to-topological transition points, we present $\mathcal{S}(L)$ with $\ln[(L/\pi)\sin(\pi/2)]$ and linearly fit their slopes in Fig.~\ref{figee} (a) and (c), respectively. For the first and second critical points, the central charge is determined as 2.0094(6) and 1.9785(6), which is consistent with the change in topological invariant $\nu$ during the phase transition. For the scaling with bipartition length $l$ in these two transition points, we plot $\mathcal{S}(l)$ against $\ln[(L/\pi)\sin(\pi l/L)]$ for $L=512a_{12}$ in Fig.~\ref{figee} (b) and (d). Slopes of these linear fittings give the central charge $c=2.0072(9)$ and $c=2.0027(3)$. The low-energy excitations on band touching points at the Fermi surface are described by CFTs~\cite{JVoit_1995}, which link the relation between the number of zero-energy pairs emerge at the phase transition points and the entanglement scaling revealed central charge. For these two reentrant transition points, the central charge $c$ derived from two different scaling types is consistent with the emergent zero-energy pairs and thus the change in winding number $\nu$. This finding suggests that the PBC ground state encodes both quantum criticality and topology.

\section{Intercell hopping $w\neq0$\label{sec4}}
When intercell hopping $w\neq 0$ $(\varepsilon<1)$ is considered, the phase diagram becomes complex, and previously employed analytical method fail to converge a concise result. Nevertheless, by leveraging the physical meaning of the momentum-space winding number, we can still numerically determine the phase boundary in the thermodynamic limit. The bulk-boundary correspondence between PBC entanglement spectrum and OBC topological edge states persists, and the universality class of various phase transitions in the reentrant sequences remains unchanged.

\begin{figure}[t!]
\centering
\includegraphics[width=0.48\textwidth]{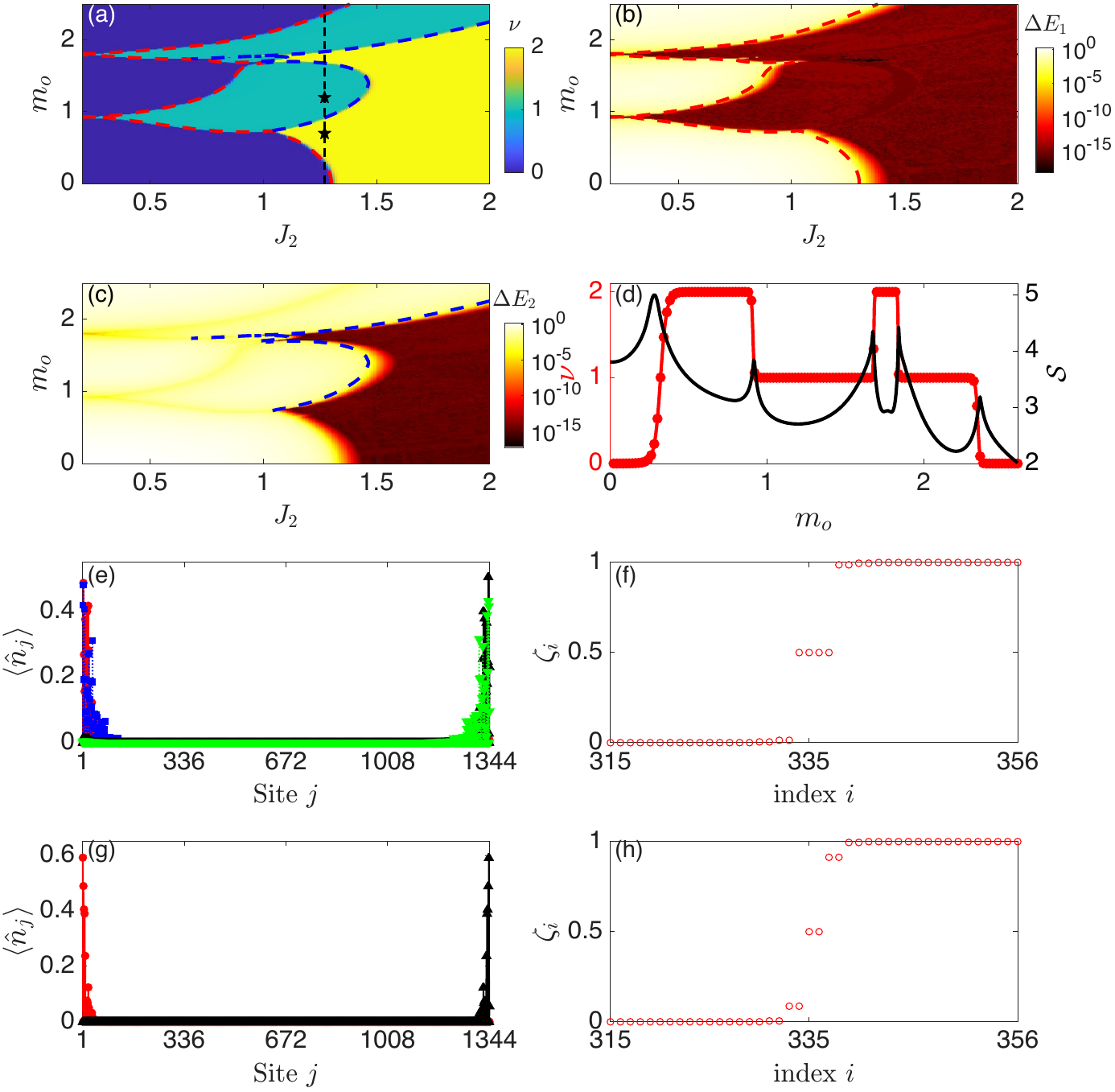}
\caption{(Color online) Characterization of reentrant physics with intercell hopping $w\neq0$. (a) The topological phase diagram for $\varepsilon=0.3$. Red and blue dashed curves are the thermodynamic limit phase boundary obtained from condition equation $q(k)=0$. The vertical black dashed line correspond to the cutting line in (d), and black pentagram marks denote the chosen parameters in (e,f) and (g,h), respectively. (b) The first energy gap $\Delta E_1$, and (c) the second energy gap $\Delta E_2$ plotted as functions of $m_o$ and $J_2$. (d) $\nu$ and $\mathcal{S}$ plotted as functions of $m_o$ for $J_2=1.27$. (e,f) Density distributions $\braket{\hat{n}_j}$, and entanglement spectrum $\zeta_i$ for $\nu=2$ with $J_2=1.27$ and $m_o=0.7$. (g,h) $\braket{\hat{n}_j}$, and $\zeta_i$ for $\nu=1$ with $J_2=1.27$ and $m_o=1.2$. Other parameters are $\varepsilon=0.3$, $A=32$, $a_1=3$ and $a_2=7$.}
\label{figApd}
\end{figure}

The phase diagram for a specific $\varepsilon=0.3$ consists of three different winding numbers, as shown in Fig.~\ref{figApd} (a), trivial regions with $\nu=0$, topological regions with $\nu=1$ and $\nu=2$. This phase diagram enables various types of reentrant sequences. Following the previously used localization length analysis, the recursive equations of the zero-energy states involve three probability parameters on different sites, rendering them analytically intractable. We then analyze the momentum-space winding number and derive the thermodynamic limit phase boundary. The $q(k)$ in Eq.~(\ref{nu_k}) traces a path in the complex plane while varying $k$ in the reduced MBZ, and the transition point corresponds to $\det q(k)$ crossing the origin. The $q(k)$ obtained from momentum-space Hamiltonian is translation invariant in terms of continuous wave vectors $k$, and effectively captures the infinite-size limit. Thus, the critical point, marked by the crossing of the origin corresponds to the phase transition point in the thermodynamic limit. The $q(k)$ is a $21\times21$ matrix for our chosen $n_{12}=21$, and the transition condition $\det q(k)=0$ is numerically solved and depicted in Fig.~\ref{figApd} (a) as red and blue dashed curves. The red dashed curve, corresponding to a phase transition from $\nu=0$ to $\nu=1$ or $\nu=2$, is consistent with the closing of the first energy gap $\Delta E_1$ presented in Fig.~\ref{figApd} (b). The blue dashed curve, marking the $\nu=1$ to $\nu=2$ topological phase boundary, matches the closing of the second energy gap $\Delta E_2$ plotted in Fig.~\ref{figApd} (c). The black dashed cutting line for $J_2=1.27$ in Fig.~\ref{figApd} (a) shows a reentrant sequence of $\nu=0\rightarrow 2 \rightarrow 1 \rightarrow 2 \rightarrow 1 \rightarrow 0$ transitions, whose OBC winding number $\nu$ along with PBC entanglement entropy $\mathcal{S}$ are shown in Fig.~\ref{figApd} (d). The exponential increase of the correlation function near the transition point is distinctly captured by the increase of entanglement entropy. The black pentagrams denote the chosen parameters for Figs.~\ref{figApd} (e,f) in the $\nu=2$ region, and Figs.~\ref{figApd} (g,h) in the $\nu=1$ region. In both cases, the bulk-boundary correspondence for PBC entanglement spectrum and OBC localized edge states persist. A four (two) degenerate zero-energy edge states corresponds to a four (two) degenerate entanglement spectrum in the $\nu=2$ ($\nu=1$) topological region.

\begin{figure}[t!]
\centering
\includegraphics[width=0.48\textwidth]{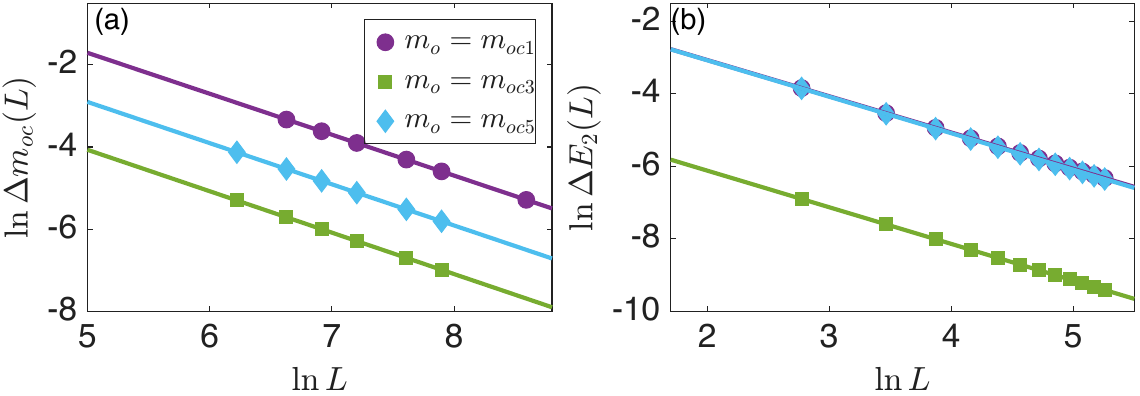}
\caption{(Color online) Critical behaviors with intercell hopping $w\neq0$. (a) $\mathrm{Log}$-$\log$ plot of system size $L$ and $\Delta m_{oc}(L)=|m_{oc}(L)-m_{oci}|$ with $m_{oc1}=0.2790(0)$ the $\nu=0\rightarrow\nu=2$ phase transition, $m_{oc3}=1.6773(8)$ the $\nu=1\rightarrow\nu=2$ phase transition, and $m_{oc5=2.3579(6)}$ the $\nu=1\rightarrow\nu=0$ phase transition. $m_{oc}^{(L)}$ is the pseudo critical point with finite-size effect revealed by the peak position of $\partial \nu / \partial m_o$ for certain $L$. (b) $\mathrm{Log}$-$\log$ plot of the second energy gap $\Delta E_2$ versus system size $L$ near critical points $m_{oc1}$, $m_{oc3}$, and $m_{oc5}$, respectively. Other parameter are $\varepsilon=0.3$, $a_1=3$, $a_2=7$, and $J_2=1.27$.}
\label{figAscal}
\end{figure}

We demonstrate through the scale-invariant analysis at three distinct phase transition points that the universality class of this reentrant phase transition model remains unchanged. The $\log$-$\log$ plot of the distance from pseudo critical points $m_{oc}^{(L)}$ to real critical point $m_{oci}$ with respect to system size $L$ for $m_{oc1}=0.2790(0)$, $m_{oc3}=1.6773(8)$, and $m_{oc5}=2.3579(6)$ is depicted in Fig.~\ref{figAscal} (a). Linear fittings for these three transitions give the critical exponents $\mu=0.9962(7)$, $\mu=1.0079(8)$, and $\mu=1.0022(0)$. While, the $\log$-$\log$ plot of the second energy gap $\Delta E_2$ near these three transition points fit linearly with dynamical critical exponents $z=0.9986(3)$, $z=1.0089(0)$, and $z=1.0049(7)$. These numerical results indicate that this moir\'e-modulated extended SSH model belongs to the same universality class as $\varepsilon=1$ case, characterized by the critical exponent $\mu\approx1$ and the dynamical critical exponent $z\approx1$. This universality class is preserved even in this complex case with intricate topological transitions.

\begin{figure}[t!]
\centering
\includegraphics[width=0.48\textwidth]{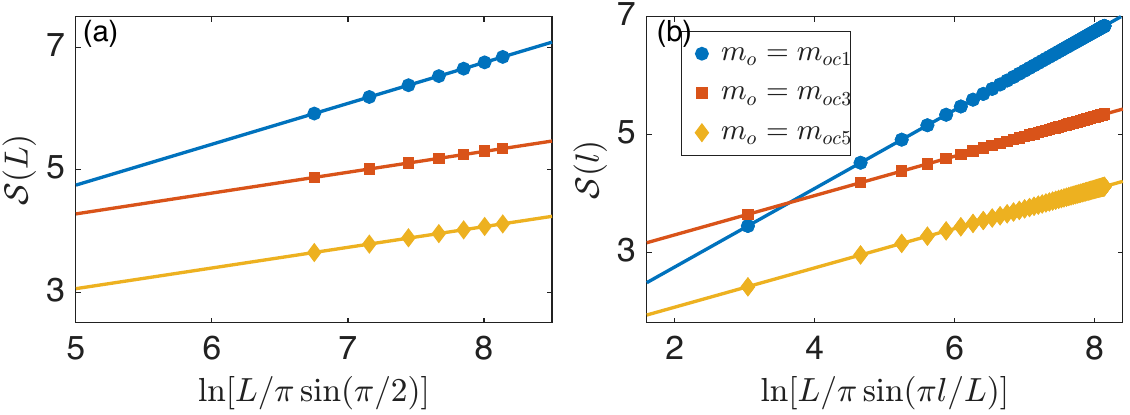}
\caption{(Color online) Entanglement scalings with intercell hopping $w\neq0$. (a) Entanglement entropy $\mathcal{S}(L)$  versus rescaled system size $L$ in three phase transition points $m_{oc1}=0.2790(0)$, $m_{oc3}=1.6773(8)$, and $m_{oc5}=2.3579(6)$.  (b) Entanglement entropy $\mathcal{S}(l)$ versus rescaled bipartition length $l$ in the same phase transition points as (a). Solid lines are linear fittings revealing the central charge $c\approx 2$, $c\approx1$, and $c\approx1$ for $m_{oc1}$, $m_{oc3}$, and $m_{oc5}$, respectively. Other parameter are $\varepsilon=0.3$, $a_1=3$, $a_2=7$, $J_2=1.6$, and $A=512$ for (b).}
\label{figAee}
\end{figure}

The bulk entanglement entropy in critical region exhibits a volume law and obeys Eq.~(\ref{eescal}). Here, we also perform two types of scaling analyses:  one with respect to the system size $L$, and the other scales against bipartition length $l$. In Fig.~\ref{figAee} (a), the half-chain entanglement entropy $\mathcal{S}(L)$ is plotted as a linear function of $\ln[L/\pi\sin(\pi/2)]$ for critical point $m_{oc1}=0.2790(0)$ (blue circles), $m_{oc3}=1.6773(8)$ (red squares), and $m_{oc5}=2.3579(6)$ (orange diamonds). Slopes of each linear fitting correspond to central charge $c=2.0005(2)$ for the $\nu=0\rightarrow\nu=2$ transition, $c=1.0193(4)$  for the $\nu=1\rightarrow\nu=2$ transition, and $c=1.0120(5)$ for the $\nu=1\rightarrow\nu=0$ transition, consistent with the change in winding number $\nu$ during each phase transition. For a certain length $L=512a_{12}$, the entanglement entropy against bipartition length $l$ is shown in Fig.~\ref{figAee} (b) as a function of $\ln[L/\pi\sin(\pi l/L)]$. Each linear fitting implies the central charge $c=2.0002(5)$, $c=1.0015(9)$, and $c=1.0012(0)$, agree with the corresponding system size $L$ fittings.

\section{\label{secf}Discussion and conclusion}
Before concluding, we briefly discuss the realization of our moir\'e-modulated extended SSH model using ultracold atoms in a synthetic momentum lattice. Implementing such models in real space is challenging as the internal dimension of the supercell, $2a_{12}$, greatly exceeds the original SSH model lattice spacing. However, this higher internal dimensionality can be achieved with trapped ultracold atoms manipulated by Raman beams~\cite{Xie2019}, where two discrete momentum states transition from $\ket{n}$ to $\ket{n+1}$ (separated by $2\hbar k$) can be coupled through a two-photon recoil process with energy $p_0=2\hbar k$. The approximated momentum lattice Hamiltonian reads~\cite{PhysRevA.92.043606}
\begin{equation}
\hat{H}_{\mathrm{R}}=\sum_n t_n (\ket{n+1}\bra{n}+\mathrm{H.c.}),
\end{equation}
where $t_n$ is the site-dependent tunneling rate determined by the Raman beams. A periodic modulation over $2a_{12}$ sites enables the emergence of the moir\'e pattern in the model couplings by setting $t_{2j-1}=v_j$, and $t_{2j}=w$. Topological phases and corresponding winding numbers $\nu$ can be probed via quench dynamics. Notably, the topological winding number is highly robust against disorder and can be extracted from the mean chiral displacement, which converges to $\nu$ in the long-time evolution~\cite{doi:10.1126/science.aat3406,Xie2019}. To realize the long-range hopping $J_2$, which corresponds to a fourth-neighbor hopping term coupling the momentum states $\ket{n}$ and $\ket{n+4}$, higher-order multi-photon Bragg transitions imparting $8\hbar k$ momentum transfer are required. Direct realizations of such long-range hopping terms remain demanding in recent experimental systems. However, as the four-photon Bragg transitions enabling next-nearest-neighbor hopping terms (coupling $\ket{n}$ and $\ket{n+2}$) have been realized in recent experiments~\cite{Li2023,Ren:23}, the realization of our proposed model appears feasible in the near future.

To summarize, we have investigated the reentrant topological phase transitions induced by moir\'e pattern in the extended SSH model. For the simplified case with $w=0$, the model Hamiltonian exists analytical solution to the topological phase transition boundary. We show the phase diagram consists with trivial region and $\nu=2$ topological region, whose boundary is analytically derived from the localization length $\Lambda$. The reentrant phase transition and the invariant of the universality class are revealed, with the reentrant phenomenon attributed to the renormalization of Hamiltonian parameter $v$ by the moir\'e strength $m_o$. We also show the correspondence between OBC topological winding number and PBC entanglement entropy, OBC zero-energy edge modes and PBC entanglement spectrum. The central charge derived from PBC entanglement entropy equals to the number of zero-energy pairs emerge during the phase transition, and thus the change in winding number. The degeneracy of PBC entanglement spectrum corresponds to the degeneracy of zero-energy edge states. Our results are extended to the $w\neq0$ case, where more intricate topological phase transition sequences occur. The analytical solution for phase boundaries is omit, but can still be derived numerically from the momentum-space winding number. By revealing these complex topological transitions, their universal characteristics in the critical region, and relation between entanglement scalings, our study establishes a comprehensive paradigm for moir\'e-modulated systems. It would also be interesting to further investigate the universality class and correspondence between topology and entanglement in interacting moir\'e systems.

\begin{acknowledgments}
G.Q.Z. acknowledges support from the National Natural Science Foundation of China (Grant No. 12104166). S.H.D. started this work on the research stay in China.
\end{acknowledgments}

\vspace{1em}
\textbf{Conflict of interest} \ The authors declare that they have no conflict of interest.
\vspace{1em}

\bibliography{reference}

\begin{thebibliography}{79}%
\makeatletter
\providecommand \@ifxundefined [1]{%
 \@ifx{#1\undefined}
}%
\providecommand \@ifnum [1]{%
 \ifnum #1\expandafter \@firstoftwo
 \else \expandafter \@secondoftwo
 \fi
}%
\providecommand \@ifx [1]{%
 \ifx #1\expandafter \@firstoftwo
 \else \expandafter \@secondoftwo
 \fi
}%
\providecommand \natexlab [1]{#1}%
\providecommand \enquote  [1]{``#1''}%
\providecommand \bibnamefont  [1]{#1}%
\providecommand \bibfnamefont [1]{#1}%
\providecommand \citenamefont [1]{#1}%
\providecommand \href@noop [0]{\@secondoftwo}%
\providecommand \href [0]{\begingroup \@sanitize@url \@href}%
\providecommand \@href[1]{\@@startlink{#1}\@@href}%
\providecommand \@@href[1]{\endgroup#1\@@endlink}%
\providecommand \@sanitize@url [0]{\catcode `\\12\catcode `\$12\catcode
  `\&12\catcode `\#12\catcode `\^12\catcode `\_12\catcode `\%12\relax}%
\providecommand \@@startlink[1]{}%
\providecommand \@@endlink[0]{}%
\providecommand \url  [0]{\begingroup\@sanitize@url \@url }%
\providecommand \@url [1]{\endgroup\@href {#1}{\urlprefix }}%
\providecommand \urlprefix  [0]{URL }%
\providecommand \Eprint [0]{\href }%
\providecommand \doibase [0]{http://dx.doi.org/}%
\providecommand \selectlanguage [0]{\@gobble}%
\providecommand \bibinfo  [0]{\@secondoftwo}%
\providecommand \bibfield  [0]{\@secondoftwo}%
\providecommand \translation [1]{[#1]}%
\providecommand \BibitemOpen [0]{}%
\providecommand \bibitemStop [0]{}%
\providecommand \bibitemNoStop [0]{.\EOS\space}%
\providecommand \EOS [0]{\spacefactor3000\relax}%
\providecommand \BibitemShut  [1]{\csname bibitem#1\endcsname}%
\let\auto@bib@innerbib\@empty
%</preamble>
\bibitem [{\citenamefont {Gon{\c{c}}alves}\ \emph {et~al.}(2024)\citenamefont
  {Gon{\c{c}}alves}, \citenamefont {Amorim}, \citenamefont {Riche},
  \citenamefont {Castro},\ and\ \citenamefont
  {Ribeiro}}]{gonccalves2024incommensurability}%
  \BibitemOpen
  \bibfield  {author} {\bibinfo {author} {\bibfnamefont {M.}~\bibnamefont
  {Gon{\c{c}}alves}}, \bibinfo {author} {\bibfnamefont {B.}~\bibnamefont
  {Amorim}}, \bibinfo {author} {\bibfnamefont {F.}~\bibnamefont {Riche}},
  \bibinfo {author} {\bibfnamefont {E.~V.}\ \bibnamefont {Castro}}, \ and\
  \bibinfo {author} {\bibfnamefont {P.}~\bibnamefont {Ribeiro}},\ }\href
  {\doibase 10.1038/s41567-024-02662-2} {\bibfield  {journal} {\bibinfo
  {journal} {Nat. Phys.}\ }\textbf {\bibinfo {volume} {20}},\ \bibinfo {pages}
  {1933} (\bibinfo {year} {2024})}\BibitemShut {NoStop}%
\bibitem [{\citenamefont {Ren}\ \emph {et~al.}(2024)\citenamefont {Ren},
  \citenamefont {Ren}, \citenamefont {Watanabe}, \citenamefont {Taniguchi},\
  and\ \citenamefont {He}}]{ren2024realizing}%
  \BibitemOpen
  \bibfield  {author} {\bibinfo {author} {\bibfnamefont {Y.-N.}\ \bibnamefont
  {Ren}}, \bibinfo {author} {\bibfnamefont {H.-Y.}\ \bibnamefont {Ren}},
  \bibinfo {author} {\bibfnamefont {K.}~\bibnamefont {Watanabe}}, \bibinfo
  {author} {\bibfnamefont {T.}~\bibnamefont {Taniguchi}}, \ and\ \bibinfo
  {author} {\bibfnamefont {L.}~\bibnamefont {He}},\ }\href {\doibase
  10.1073/pnas.2405582121} {\bibfield  {journal} {\bibinfo  {journal} {Proc.
  Natl. Acad. Sci.}\ }\textbf {\bibinfo {volume} {121}},\ \bibinfo {pages}
  {e2405582121} (\bibinfo {year} {2024})}\BibitemShut {NoStop}%
\bibitem [{\citenamefont {Zhou}\ \emph {et~al.}(2024)\citenamefont {Zhou},
  \citenamefont {Chen}, \citenamefont {Chen}, \citenamefont {Hu}, \citenamefont
  {Lyu}, \citenamefont {Xu}, \citenamefont {Lou}, \citenamefont {Shen},
  \citenamefont {Ma}, \citenamefont {Wu}, \citenamefont {Xie}, \citenamefont
  {Zhang}, \citenamefont {L\"u}, \citenamefont {Luo}, \citenamefont {Liang},
  \citenamefont {Xian}, \citenamefont {Zhang},\ and\ \citenamefont
  {Shi}}]{PhysRevB.109.045105}%
  \BibitemOpen
  \bibfield  {author} {\bibinfo {author} {\bibfnamefont {X.}~\bibnamefont
  {Zhou}}, \bibinfo {author} {\bibfnamefont {Y.}~\bibnamefont {Chen}}, \bibinfo
  {author} {\bibfnamefont {J.}~\bibnamefont {Chen}}, \bibinfo {author}
  {\bibfnamefont {C.}~\bibnamefont {Hu}}, \bibinfo {author} {\bibfnamefont
  {B.}~\bibnamefont {Lyu}}, \bibinfo {author} {\bibfnamefont {K.}~\bibnamefont
  {Xu}}, \bibinfo {author} {\bibfnamefont {S.}~\bibnamefont {Lou}}, \bibinfo
  {author} {\bibfnamefont {P.}~\bibnamefont {Shen}}, \bibinfo {author}
  {\bibfnamefont {S.}~\bibnamefont {Ma}}, \bibinfo {author} {\bibfnamefont
  {Z.}~\bibnamefont {Wu}}, \bibinfo {author} {\bibfnamefont {Y.}~\bibnamefont
  {Xie}}, \bibinfo {author} {\bibfnamefont {Z.}~\bibnamefont {Zhang}}, \bibinfo
  {author} {\bibfnamefont {Z.}~\bibnamefont {L\"u}}, \bibinfo {author}
  {\bibfnamefont {W.}~\bibnamefont {Luo}}, \bibinfo {author} {\bibfnamefont
  {Q.}~\bibnamefont {Liang}}, \bibinfo {author} {\bibfnamefont
  {L.}~\bibnamefont {Xian}}, \bibinfo {author} {\bibfnamefont {G.}~\bibnamefont
  {Zhang}}, \ and\ \bibinfo {author} {\bibfnamefont {Z.}~\bibnamefont {Shi}},\
  }\href {\doibase 10.1103/PhysRevB.109.045105} {\bibfield  {journal} {\bibinfo
   {journal} {Phys. Rev. B}\ }\textbf {\bibinfo {volume} {109}},\ \bibinfo
  {pages} {045105} (\bibinfo {year} {2024})}\BibitemShut {NoStop}%
\bibitem [{\citenamefont {Zhang}\ \emph {et~al.}(2025)\citenamefont {Zhang},
  \citenamefont {Tang}, \citenamefont {Quezada}, \citenamefont {Dong},\ and\
  \citenamefont {Zhang}}]{zhang2025reentrant}%
  \BibitemOpen
  \bibfield  {author} {\bibinfo {author} {\bibfnamefont {G.-Q.}\ \bibnamefont
  {Zhang}}, \bibinfo {author} {\bibfnamefont {L.-Z.}\ \bibnamefont {Tang}},
  \bibinfo {author} {\bibfnamefont {L.}~\bibnamefont {Quezada}}, \bibinfo
  {author} {\bibfnamefont {S.-H.}\ \bibnamefont {Dong}}, \ and\ \bibinfo
  {author} {\bibfnamefont {D.-W.}\ \bibnamefont {Zhang}},\ }\href {\doibase
  10.1038/s42005-025-02197-9} {\bibfield  {journal} {\bibinfo  {journal}
  {Commun. Phys.}\ }\textbf {\bibinfo {volume} {8}},\ \bibinfo {pages} {275}
  (\bibinfo {year} {2025})}\BibitemShut {NoStop}%
\bibitem [{\citenamefont {Cao}\ \emph {et~al.}(2018{\natexlab{a}})\citenamefont
  {Cao}, \citenamefont {Fatemi}, \citenamefont {Demir}, \citenamefont {Fang},
  \citenamefont {Tomarken}, \citenamefont {Luo}, \citenamefont
  {Sanchez-Yamagishi}, \citenamefont {Watanabe}, \citenamefont {Taniguchi},
  \citenamefont {Kaxiras}, \citenamefont {Ashoori},\ and\ \citenamefont
  {Jarillo-Herrero}}]{cao2018correlated}%
  \BibitemOpen
  \bibfield  {author} {\bibinfo {author} {\bibfnamefont {Y.}~\bibnamefont
  {Cao}}, \bibinfo {author} {\bibfnamefont {V.}~\bibnamefont {Fatemi}},
  \bibinfo {author} {\bibfnamefont {A.}~\bibnamefont {Demir}}, \bibinfo
  {author} {\bibfnamefont {S.}~\bibnamefont {Fang}}, \bibinfo {author}
  {\bibfnamefont {S.~L.}\ \bibnamefont {Tomarken}}, \bibinfo {author}
  {\bibfnamefont {J.~Y.}\ \bibnamefont {Luo}}, \bibinfo {author} {\bibfnamefont
  {J.~D.}\ \bibnamefont {Sanchez-Yamagishi}}, \bibinfo {author} {\bibfnamefont
  {K.}~\bibnamefont {Watanabe}}, \bibinfo {author} {\bibfnamefont
  {T.}~\bibnamefont {Taniguchi}}, \bibinfo {author} {\bibfnamefont
  {E.}~\bibnamefont {Kaxiras}}, \bibinfo {author} {\bibfnamefont {R.~C.}\
  \bibnamefont {Ashoori}}, \ and\ \bibinfo {author} {\bibfnamefont
  {P.}~\bibnamefont {Jarillo-Herrero}},\ }\href {\doibase 10.1038/nature26154}
  {\bibfield  {journal} {\bibinfo  {journal} {Nature}\ }\textbf {\bibinfo
  {volume} {556}},\ \bibinfo {pages} {80} (\bibinfo {year}
  {2018}{\natexlab{a}})}\BibitemShut {NoStop}%
\bibitem [{\citenamefont {Balents}\ \emph {et~al.}(2020)\citenamefont
  {Balents}, \citenamefont {Dean}, \citenamefont {Efetov},\ and\ \citenamefont
  {Young}}]{balents2020superconductivity}%
  \BibitemOpen
  \bibfield  {author} {\bibinfo {author} {\bibfnamefont {L.}~\bibnamefont
  {Balents}}, \bibinfo {author} {\bibfnamefont {C.~R.}\ \bibnamefont {Dean}},
  \bibinfo {author} {\bibfnamefont {D.~K.}\ \bibnamefont {Efetov}}, \ and\
  \bibinfo {author} {\bibfnamefont {A.~F.}\ \bibnamefont {Young}},\ }\href
  {\doibase 10.1038/s41567-020-0906-9} {\bibfield  {journal} {\bibinfo
  {journal} {Nat. Phys.}\ }\textbf {\bibinfo {volume} {16}},\ \bibinfo {pages}
  {725} (\bibinfo {year} {2020})}\BibitemShut {NoStop}%
\bibitem [{\citenamefont {Hu}\ \emph {et~al.}(2023)\citenamefont {Hu},
  \citenamefont {Zhang}, \citenamefont {Hu}, \citenamefont {Sun}, \citenamefont
  {Wang}, \citenamefont {Lin},\ and\ \citenamefont {Li}}]{Hu2023}%
  \BibitemOpen
  \bibfield  {author} {\bibinfo {author} {\bibfnamefont {J.}~\bibnamefont
  {Hu}}, \bibinfo {author} {\bibfnamefont {X.}~\bibnamefont {Zhang}}, \bibinfo
  {author} {\bibfnamefont {C.}~\bibnamefont {Hu}}, \bibinfo {author}
  {\bibfnamefont {J.}~\bibnamefont {Sun}}, \bibinfo {author} {\bibfnamefont
  {X.}~\bibnamefont {Wang}}, \bibinfo {author} {\bibfnamefont {H.-Q.}\
  \bibnamefont {Lin}}, \ and\ \bibinfo {author} {\bibfnamefont
  {G.}~\bibnamefont {Li}},\ }\href {\doibase 10.1038/s42005-023-01292-z}
  {\bibfield  {journal} {\bibinfo  {journal} {Commun. Phys.}\ }\textbf
  {\bibinfo {volume} {6}},\ \bibinfo {pages} {172} (\bibinfo {year}
  {2023})}\BibitemShut {NoStop}%
\bibitem [{\citenamefont {Cr{\'e}pel}\ \emph {et~al.}(2024)\citenamefont
  {Cr{\'e}pel}, \citenamefont {Regnault},\ and\ \citenamefont
  {Queiroz}}]{Crepel2024}%
  \BibitemOpen
  \bibfield  {author} {\bibinfo {author} {\bibfnamefont {V.}~\bibnamefont
  {Cr{\'e}pel}}, \bibinfo {author} {\bibfnamefont {N.}~\bibnamefont
  {Regnault}}, \ and\ \bibinfo {author} {\bibfnamefont {R.}~\bibnamefont
  {Queiroz}},\ }\href {\doibase 10.1038/s42005-024-01641-6} {\bibfield
  {journal} {\bibinfo  {journal} {Commun. Phys.}\ }\textbf {\bibinfo {volume}
  {7}},\ \bibinfo {pages} {146} (\bibinfo {year} {2024})}\BibitemShut {NoStop}%
\bibitem [{\citenamefont {Zhang}\ \emph
  {et~al.}(2020{\natexlab{a}})\citenamefont {Zhang}, \citenamefont {Zhang},
  \citenamefont {Wu}, \citenamefont {Wang}, \citenamefont {Gogna},
  \citenamefont {Hou}, \citenamefont {Watanabe}, \citenamefont {Taniguchi},
  \citenamefont {Kulkarni}, \citenamefont {Kuo}, \citenamefont {Forrest},\ and\
  \citenamefont {Deng}}]{zhang2020twist}%
  \BibitemOpen
  \bibfield  {author} {\bibinfo {author} {\bibfnamefont {L.}~\bibnamefont
  {Zhang}}, \bibinfo {author} {\bibfnamefont {Z.}~\bibnamefont {Zhang}},
  \bibinfo {author} {\bibfnamefont {F.}~\bibnamefont {Wu}}, \bibinfo {author}
  {\bibfnamefont {D.}~\bibnamefont {Wang}}, \bibinfo {author} {\bibfnamefont
  {R.}~\bibnamefont {Gogna}}, \bibinfo {author} {\bibfnamefont
  {S.}~\bibnamefont {Hou}}, \bibinfo {author} {\bibfnamefont {K.}~\bibnamefont
  {Watanabe}}, \bibinfo {author} {\bibfnamefont {T.}~\bibnamefont {Taniguchi}},
  \bibinfo {author} {\bibfnamefont {K.}~\bibnamefont {Kulkarni}}, \bibinfo
  {author} {\bibfnamefont {T.}~\bibnamefont {Kuo}}, \bibinfo {author}
  {\bibfnamefont {S.~R.}\ \bibnamefont {Forrest}}, \ and\ \bibinfo {author}
  {\bibfnamefont {H.}~\bibnamefont {Deng}},\ }\href {\doibase
  10.1038/s41467-020-19466-6} {\bibfield  {journal} {\bibinfo  {journal} {Nat.
  Commun.}\ }\textbf {\bibinfo {volume} {11}},\ \bibinfo {pages} {5888}
  (\bibinfo {year} {2020}{\natexlab{a}})}\BibitemShut {NoStop}%
\bibitem [{\citenamefont {Cui}\ \emph {et~al.}(2024)\citenamefont {Cui},
  \citenamefont {Le}, \citenamefont {Zhang}, \citenamefont {Wu}, \citenamefont
  {Hu},\ and\ \citenamefont {Chiu}}]{Cui2024}%
  \BibitemOpen
  \bibfield  {author} {\bibinfo {author} {\bibfnamefont {F.}~\bibnamefont
  {Cui}}, \bibinfo {author} {\bibfnamefont {C.}~\bibnamefont {Le}}, \bibinfo
  {author} {\bibfnamefont {Q.}~\bibnamefont {Zhang}}, \bibinfo {author}
  {\bibfnamefont {X.}~\bibnamefont {Wu}}, \bibinfo {author} {\bibfnamefont
  {J.}~\bibnamefont {Hu}}, \ and\ \bibinfo {author} {\bibfnamefont {C.-K.}\
  \bibnamefont {Chiu}},\ }\href {\doibase 10.1007/s11433-024-2410-9} {\bibfield
   {journal} {\bibinfo  {journal} {Sci. China-Phys. Mech. Astron.}\ }\textbf
  {\bibinfo {volume} {67}},\ \bibinfo {pages} {297012} (\bibinfo {year}
  {2024})}\BibitemShut {NoStop}%
\bibitem [{\citenamefont {Dean}\ \emph {et~al.}(2013)\citenamefont {Dean},
  \citenamefont {Wang}, \citenamefont {Maher}, \citenamefont {Forsythe},
  \citenamefont {Ghahari}, \citenamefont {Gao}, \citenamefont {Katoch},
  \citenamefont {Ishigami}, \citenamefont {Moon}, \citenamefont {Koshino},
  \citenamefont {Taniguchi}, \citenamefont {Watanabe}, \citenamefont {Shepard},
  \citenamefont {Hone},\ and\ \citenamefont {Kim}}]{dean2013hofstadter}%
  \BibitemOpen
  \bibfield  {author} {\bibinfo {author} {\bibfnamefont {C.~R.}\ \bibnamefont
  {Dean}}, \bibinfo {author} {\bibfnamefont {L.}~\bibnamefont {Wang}}, \bibinfo
  {author} {\bibfnamefont {P.}~\bibnamefont {Maher}}, \bibinfo {author}
  {\bibfnamefont {C.}~\bibnamefont {Forsythe}}, \bibinfo {author}
  {\bibfnamefont {F.}~\bibnamefont {Ghahari}}, \bibinfo {author} {\bibfnamefont
  {Y.}~\bibnamefont {Gao}}, \bibinfo {author} {\bibfnamefont {J.}~\bibnamefont
  {Katoch}}, \bibinfo {author} {\bibfnamefont {M.}~\bibnamefont {Ishigami}},
  \bibinfo {author} {\bibfnamefont {P.}~\bibnamefont {Moon}}, \bibinfo {author}
  {\bibfnamefont {M.}~\bibnamefont {Koshino}}, \bibinfo {author} {\bibfnamefont
  {T.}~\bibnamefont {Taniguchi}}, \bibinfo {author} {\bibfnamefont
  {K.}~\bibnamefont {Watanabe}}, \bibinfo {author} {\bibfnamefont {K.~L.}\
  \bibnamefont {Shepard}}, \bibinfo {author} {\bibfnamefont {J.}~\bibnamefont
  {Hone}}, \ and\ \bibinfo {author} {\bibfnamefont {P.}~\bibnamefont {Kim}},\
  }\href {\doibase 10.1038/nature12186} {\bibfield  {journal} {\bibinfo
  {journal} {Nature}\ }\textbf {\bibinfo {volume} {497}},\ \bibinfo {pages}
  {598} (\bibinfo {year} {2013})}\BibitemShut {NoStop}%
\bibitem [{\citenamefont {Chittari}\ \emph {et~al.}(2019)\citenamefont
  {Chittari}, \citenamefont {Chen}, \citenamefont {Zhang}, \citenamefont
  {Wang},\ and\ \citenamefont {Jung}}]{PhysRevLett.122.016401}%
  \BibitemOpen
  \bibfield  {author} {\bibinfo {author} {\bibfnamefont {B.~L.}\ \bibnamefont
  {Chittari}}, \bibinfo {author} {\bibfnamefont {G.}~\bibnamefont {Chen}},
  \bibinfo {author} {\bibfnamefont {Y.}~\bibnamefont {Zhang}}, \bibinfo
  {author} {\bibfnamefont {F.}~\bibnamefont {Wang}}, \ and\ \bibinfo {author}
  {\bibfnamefont {J.}~\bibnamefont {Jung}},\ }\href {\doibase
  10.1103/PhysRevLett.122.016401} {\bibfield  {journal} {\bibinfo  {journal}
  {Phys. Rev. Lett.}\ }\textbf {\bibinfo {volume} {122}},\ \bibinfo {pages}
  {016401} (\bibinfo {year} {2019})}\BibitemShut {NoStop}%
\bibitem [{\citenamefont {Serlin}\ \emph {et~al.}(2020)\citenamefont {Serlin},
  \citenamefont {Tschirhart}, \citenamefont {Polshyn}, \citenamefont {Zhang},
  \citenamefont {Zhu}, \citenamefont {Watanabe}, \citenamefont {Taniguchi},
  \citenamefont {Balents},\ and\ \citenamefont
  {Young}}]{doi:10.1126/science.aay5533}%
  \BibitemOpen
  \bibfield  {author} {\bibinfo {author} {\bibfnamefont {M.}~\bibnamefont
  {Serlin}}, \bibinfo {author} {\bibfnamefont {C.~L.}\ \bibnamefont
  {Tschirhart}}, \bibinfo {author} {\bibfnamefont {H.}~\bibnamefont {Polshyn}},
  \bibinfo {author} {\bibfnamefont {Y.}~\bibnamefont {Zhang}}, \bibinfo
  {author} {\bibfnamefont {J.}~\bibnamefont {Zhu}}, \bibinfo {author}
  {\bibfnamefont {K.}~\bibnamefont {Watanabe}}, \bibinfo {author}
  {\bibfnamefont {T.}~\bibnamefont {Taniguchi}}, \bibinfo {author}
  {\bibfnamefont {L.}~\bibnamefont {Balents}}, \ and\ \bibinfo {author}
  {\bibfnamefont {A.~F.}\ \bibnamefont {Young}},\ }\href {\doibase
  10.1126/science.aay5533} {\bibfield  {journal} {\bibinfo  {journal}
  {Science}\ }\textbf {\bibinfo {volume} {367}},\ \bibinfo {pages} {900}
  (\bibinfo {year} {2020})}\BibitemShut {NoStop}%
\bibitem [{\citenamefont {Cao}\ \emph {et~al.}(2018{\natexlab{b}})\citenamefont
  {Cao}, \citenamefont {Fatemi}, \citenamefont {Fang}, \citenamefont
  {Watanabe}, \citenamefont {Taniguchi}, \citenamefont {Kaxiras},\ and\
  \citenamefont {Jarillo-Herrero}}]{cao2018unconventional}%
  \BibitemOpen
  \bibfield  {author} {\bibinfo {author} {\bibfnamefont {Y.}~\bibnamefont
  {Cao}}, \bibinfo {author} {\bibfnamefont {V.}~\bibnamefont {Fatemi}},
  \bibinfo {author} {\bibfnamefont {S.}~\bibnamefont {Fang}}, \bibinfo {author}
  {\bibfnamefont {K.}~\bibnamefont {Watanabe}}, \bibinfo {author}
  {\bibfnamefont {T.}~\bibnamefont {Taniguchi}}, \bibinfo {author}
  {\bibfnamefont {E.}~\bibnamefont {Kaxiras}}, \ and\ \bibinfo {author}
  {\bibfnamefont {P.}~\bibnamefont {Jarillo-Herrero}},\ }\href {\doibase
  10.1038/nature26160} {\bibfield  {journal} {\bibinfo  {journal} {Nature}\
  }\textbf {\bibinfo {volume} {556}},\ \bibinfo {pages} {43} (\bibinfo {year}
  {2018}{\natexlab{b}})}\BibitemShut {NoStop}%
\bibitem [{\citenamefont {Xie}\ and\ \citenamefont
  {Law}(2023)}]{PhysRevLett.131.016001}%
  \BibitemOpen
  \bibfield  {author} {\bibinfo {author} {\bibfnamefont {Y.-M.}\ \bibnamefont
  {Xie}}\ and\ \bibinfo {author} {\bibfnamefont {K.~T.}\ \bibnamefont {Law}},\
  }\href {\doibase 10.1103/PhysRevLett.131.016001} {\bibfield  {journal}
  {\bibinfo  {journal} {Phys. Rev. Lett.}\ }\textbf {\bibinfo {volume} {131}},\
  \bibinfo {pages} {016001} (\bibinfo {year} {2023})}\BibitemShut {NoStop}%
\bibitem [{\citenamefont {Nuckolls}\ \emph {et~al.}(2020)\citenamefont
  {Nuckolls}, \citenamefont {Oh}, \citenamefont {Wong}, \citenamefont {Lian},
  \citenamefont {Watanabe}, \citenamefont {Taniguchi}, \citenamefont
  {Bernevig},\ and\ \citenamefont {Yazdani}}]{nuckolls2020strongly}%
  \BibitemOpen
  \bibfield  {author} {\bibinfo {author} {\bibfnamefont {K.~P.}\ \bibnamefont
  {Nuckolls}}, \bibinfo {author} {\bibfnamefont {M.}~\bibnamefont {Oh}},
  \bibinfo {author} {\bibfnamefont {D.}~\bibnamefont {Wong}}, \bibinfo {author}
  {\bibfnamefont {B.}~\bibnamefont {Lian}}, \bibinfo {author} {\bibfnamefont
  {K.}~\bibnamefont {Watanabe}}, \bibinfo {author} {\bibfnamefont
  {T.}~\bibnamefont {Taniguchi}}, \bibinfo {author} {\bibfnamefont {B.~A.}\
  \bibnamefont {Bernevig}}, \ and\ \bibinfo {author} {\bibfnamefont
  {A.}~\bibnamefont {Yazdani}},\ }\href {\doibase 10.1038/s41586-020-3028-8}
  {\bibfield  {journal} {\bibinfo  {journal} {Nature}\ }\textbf {\bibinfo
  {volume} {588}},\ \bibinfo {pages} {610} (\bibinfo {year}
  {2020})}\BibitemShut {NoStop}%
\bibitem [{\citenamefont {Brei\o{}}\ and\ \citenamefont
  {Andersen}(2023)}]{PhysRevB.107.165114}%
  \BibitemOpen
  \bibfield  {author} {\bibinfo {author} {\bibfnamefont {C.~N.}\ \bibnamefont
  {Brei\o{}}}\ and\ \bibinfo {author} {\bibfnamefont {B.~M.}\ \bibnamefont
  {Andersen}},\ }\href {\doibase 10.1103/PhysRevB.107.165114} {\bibfield
  {journal} {\bibinfo  {journal} {Phys. Rev. B}\ }\textbf {\bibinfo {volume}
  {107}},\ \bibinfo {pages} {165114} (\bibinfo {year} {2023})}\BibitemShut
  {NoStop}%
\bibitem [{\citenamefont {Wang}\ \emph {et~al.}(2024)\citenamefont {Wang},
  \citenamefont {Zhang}, \citenamefont {Liu}, \citenamefont {He}, \citenamefont
  {Xu}, \citenamefont {Ran}, \citenamefont {Cao},\ and\ \citenamefont
  {Xiao}}]{PhysRevLett.132.036501}%
  \BibitemOpen
  \bibfield  {author} {\bibinfo {author} {\bibfnamefont {C.}~\bibnamefont
  {Wang}}, \bibinfo {author} {\bibfnamefont {X.-W.}\ \bibnamefont {Zhang}},
  \bibinfo {author} {\bibfnamefont {X.}~\bibnamefont {Liu}}, \bibinfo {author}
  {\bibfnamefont {Y.}~\bibnamefont {He}}, \bibinfo {author} {\bibfnamefont
  {X.}~\bibnamefont {Xu}}, \bibinfo {author} {\bibfnamefont {Y.}~\bibnamefont
  {Ran}}, \bibinfo {author} {\bibfnamefont {T.}~\bibnamefont {Cao}}, \ and\
  \bibinfo {author} {\bibfnamefont {D.}~\bibnamefont {Xiao}},\ }\href {\doibase
  10.1103/PhysRevLett.132.036501} {\bibfield  {journal} {\bibinfo  {journal}
  {Phys. Rev. Lett.}\ }\textbf {\bibinfo {volume} {132}},\ \bibinfo {pages}
  {036501} (\bibinfo {year} {2024})}\BibitemShut {NoStop}%
\bibitem [{\citenamefont {Neupert}\ \emph {et~al.}(2011)\citenamefont
  {Neupert}, \citenamefont {Santos}, \citenamefont {Chamon},\ and\
  \citenamefont {Mudry}}]{PhysRevLett.106.236804}%
  \BibitemOpen
  \bibfield  {author} {\bibinfo {author} {\bibfnamefont {T.}~\bibnamefont
  {Neupert}}, \bibinfo {author} {\bibfnamefont {L.}~\bibnamefont {Santos}},
  \bibinfo {author} {\bibfnamefont {C.}~\bibnamefont {Chamon}}, \ and\ \bibinfo
  {author} {\bibfnamefont {C.}~\bibnamefont {Mudry}},\ }\href {\doibase
  10.1103/PhysRevLett.106.236804} {\bibfield  {journal} {\bibinfo  {journal}
  {Phys. Rev. Lett.}\ }\textbf {\bibinfo {volume} {106}},\ \bibinfo {pages}
  {236804} (\bibinfo {year} {2011})}\BibitemShut {NoStop}%
\bibitem [{\citenamefont {Xu}\ \emph {et~al.}(2023)\citenamefont {Xu},
  \citenamefont {Sun}, \citenamefont {Jia}, \citenamefont {Liu}, \citenamefont
  {Xu}, \citenamefont {Li}, \citenamefont {Gu}, \citenamefont {Watanabe},
  \citenamefont {Taniguchi}, \citenamefont {Tong}, \citenamefont {Jia},
  \citenamefont {Shi}, \citenamefont {Jiang}, \citenamefont {Zhang},
  \citenamefont {Liu},\ and\ \citenamefont {Li}}]{PhysRevX.13.031037}%
  \BibitemOpen
  \bibfield  {author} {\bibinfo {author} {\bibfnamefont {F.}~\bibnamefont
  {Xu}}, \bibinfo {author} {\bibfnamefont {Z.}~\bibnamefont {Sun}}, \bibinfo
  {author} {\bibfnamefont {T.}~\bibnamefont {Jia}}, \bibinfo {author}
  {\bibfnamefont {C.}~\bibnamefont {Liu}}, \bibinfo {author} {\bibfnamefont
  {C.}~\bibnamefont {Xu}}, \bibinfo {author} {\bibfnamefont {C.}~\bibnamefont
  {Li}}, \bibinfo {author} {\bibfnamefont {Y.}~\bibnamefont {Gu}}, \bibinfo
  {author} {\bibfnamefont {K.}~\bibnamefont {Watanabe}}, \bibinfo {author}
  {\bibfnamefont {T.}~\bibnamefont {Taniguchi}}, \bibinfo {author}
  {\bibfnamefont {B.}~\bibnamefont {Tong}}, \bibinfo {author} {\bibfnamefont
  {J.}~\bibnamefont {Jia}}, \bibinfo {author} {\bibfnamefont {Z.}~\bibnamefont
  {Shi}}, \bibinfo {author} {\bibfnamefont {S.}~\bibnamefont {Jiang}}, \bibinfo
  {author} {\bibfnamefont {Y.}~\bibnamefont {Zhang}}, \bibinfo {author}
  {\bibfnamefont {X.}~\bibnamefont {Liu}}, \ and\ \bibinfo {author}
  {\bibfnamefont {T.}~\bibnamefont {Li}},\ }\href {\doibase
  10.1103/PhysRevX.13.031037} {\bibfield  {journal} {\bibinfo  {journal} {Phys.
  Rev. X}\ }\textbf {\bibinfo {volume} {13}},\ \bibinfo {pages} {031037}
  (\bibinfo {year} {2023})}\BibitemShut {NoStop}%
\bibitem [{\citenamefont {Cai}\ \emph {et~al.}(2023)\citenamefont {Cai},
  \citenamefont {Anderson}, \citenamefont {Wang}, \citenamefont {Zhang},
  \citenamefont {Liu}, \citenamefont {Holtzmann}, \citenamefont {Zhang},
  \citenamefont {Fan}, \citenamefont {Taniguchi}, \citenamefont {Watanabe},
  \citenamefont {Ran}, \citenamefont {Cao}, \citenamefont {Fu}, \citenamefont
  {Xiao}, \citenamefont {Yao},\ and\ \citenamefont {Xu}}]{Cai2023}%
  \BibitemOpen
  \bibfield  {author} {\bibinfo {author} {\bibfnamefont {J.}~\bibnamefont
  {Cai}}, \bibinfo {author} {\bibfnamefont {E.}~\bibnamefont {Anderson}},
  \bibinfo {author} {\bibfnamefont {C.}~\bibnamefont {Wang}}, \bibinfo {author}
  {\bibfnamefont {X.}~\bibnamefont {Zhang}}, \bibinfo {author} {\bibfnamefont
  {X.}~\bibnamefont {Liu}}, \bibinfo {author} {\bibfnamefont {W.}~\bibnamefont
  {Holtzmann}}, \bibinfo {author} {\bibfnamefont {Y.}~\bibnamefont {Zhang}},
  \bibinfo {author} {\bibfnamefont {F.}~\bibnamefont {Fan}}, \bibinfo {author}
  {\bibfnamefont {T.}~\bibnamefont {Taniguchi}}, \bibinfo {author}
  {\bibfnamefont {K.}~\bibnamefont {Watanabe}}, \bibinfo {author}
  {\bibfnamefont {Y.}~\bibnamefont {Ran}}, \bibinfo {author} {\bibfnamefont
  {T.}~\bibnamefont {Cao}}, \bibinfo {author} {\bibfnamefont {L.}~\bibnamefont
  {Fu}}, \bibinfo {author} {\bibfnamefont {D.}~\bibnamefont {Xiao}}, \bibinfo
  {author} {\bibfnamefont {W.}~\bibnamefont {Yao}}, \ and\ \bibinfo {author}
  {\bibfnamefont {X.}~\bibnamefont {Xu}},\ }\href {\doibase
  10.1038/s41586-023-06289-w} {\bibfield  {journal} {\bibinfo  {journal}
  {Nature}\ }\textbf {\bibinfo {volume} {622}},\ \bibinfo {pages} {63}
  (\bibinfo {year} {2023})}\BibitemShut {NoStop}%
\bibitem [{\citenamefont {Huang}\ \emph {et~al.}(2024)\citenamefont {Huang},
  \citenamefont {Li}, \citenamefont {Das~Sarma},\ and\ \citenamefont
  {Zhang}}]{PhysRevB.110.115146}%
  \BibitemOpen
  \bibfield  {author} {\bibinfo {author} {\bibfnamefont {K.}~\bibnamefont
  {Huang}}, \bibinfo {author} {\bibfnamefont {X.}~\bibnamefont {Li}}, \bibinfo
  {author} {\bibfnamefont {S.}~\bibnamefont {Das~Sarma}}, \ and\ \bibinfo
  {author} {\bibfnamefont {F.}~\bibnamefont {Zhang}},\ }\href {\doibase
  10.1103/PhysRevB.110.115146} {\bibfield  {journal} {\bibinfo  {journal}
  {Phys. Rev. B}\ }\textbf {\bibinfo {volume} {110}},\ \bibinfo {pages}
  {115146} (\bibinfo {year} {2024})}\BibitemShut {NoStop}%
\bibitem [{\citenamefont {Lu}\ \emph {et~al.}(2024)\citenamefont {Lu},
  \citenamefont {Han}, \citenamefont {Yao}, \citenamefont {Reddy},
  \citenamefont {Yang}, \citenamefont {Seo}, \citenamefont {Watanabe},
  \citenamefont {Taniguchi}, \citenamefont {Fu},\ and\ \citenamefont
  {Ju}}]{Lu2024}%
  \BibitemOpen
  \bibfield  {author} {\bibinfo {author} {\bibfnamefont {Z.}~\bibnamefont
  {Lu}}, \bibinfo {author} {\bibfnamefont {T.}~\bibnamefont {Han}}, \bibinfo
  {author} {\bibfnamefont {Y.}~\bibnamefont {Yao}}, \bibinfo {author}
  {\bibfnamefont {A.~P.}\ \bibnamefont {Reddy}}, \bibinfo {author}
  {\bibfnamefont {J.}~\bibnamefont {Yang}}, \bibinfo {author} {\bibfnamefont
  {J.}~\bibnamefont {Seo}}, \bibinfo {author} {\bibfnamefont {K.}~\bibnamefont
  {Watanabe}}, \bibinfo {author} {\bibfnamefont {T.}~\bibnamefont {Taniguchi}},
  \bibinfo {author} {\bibfnamefont {L.}~\bibnamefont {Fu}}, \ and\ \bibinfo
  {author} {\bibfnamefont {L.}~\bibnamefont {Ju}},\ }\href {\doibase
  10.1038/s41586-023-07010-7} {\bibfield  {journal} {\bibinfo  {journal}
  {Nature}\ }\textbf {\bibinfo {volume} {626}},\ \bibinfo {pages} {759}
  (\bibinfo {year} {2024})}\BibitemShut {NoStop}%
\bibitem [{\citenamefont {Zhao}\ \emph {et~al.}(2025)\citenamefont {Zhao},
  \citenamefont {Liu}, \citenamefont {Zhang}, \citenamefont {Zhang},
  \citenamefont {Feng}, \citenamefont {Wang}, \citenamefont {Lai},
  \citenamefont {Chang}, \citenamefont {Yang},\ and\ \citenamefont
  {Gao}}]{Zhao2025}%
  \BibitemOpen
  \bibfield  {author} {\bibinfo {author} {\bibfnamefont {J.}~\bibnamefont
  {Zhao}}, \bibinfo {author} {\bibfnamefont {L.}~\bibnamefont {Liu}}, \bibinfo
  {author} {\bibfnamefont {Y.}~\bibnamefont {Zhang}}, \bibinfo {author}
  {\bibfnamefont {H.}~\bibnamefont {Zhang}}, \bibinfo {author} {\bibfnamefont
  {Z.}~\bibnamefont {Feng}}, \bibinfo {author} {\bibfnamefont {C.}~\bibnamefont
  {Wang}}, \bibinfo {author} {\bibfnamefont {S.}~\bibnamefont {Lai}}, \bibinfo
  {author} {\bibfnamefont {G.}~\bibnamefont {Chang}}, \bibinfo {author}
  {\bibfnamefont {B.}~\bibnamefont {Yang}}, \ and\ \bibinfo {author}
  {\bibfnamefont {W.}~\bibnamefont {Gao}},\ }\href {\doibase
  10.1021/acsnano.5c01598} {\bibfield  {journal} {\bibinfo  {journal} {ACS
  Nano}\ }\textbf {\bibinfo {volume} {19}},\ \bibinfo {pages} {19509} (\bibinfo
  {year} {2025})}\BibitemShut {NoStop}%
\bibitem [{\citenamefont {Koshino}\ \emph {et~al.}(2015)\citenamefont
  {Koshino}, \citenamefont {Moon},\ and\ \citenamefont
  {Son}}]{PhysRevB.91.035405}%
  \BibitemOpen
  \bibfield  {author} {\bibinfo {author} {\bibfnamefont {M.}~\bibnamefont
  {Koshino}}, \bibinfo {author} {\bibfnamefont {P.}~\bibnamefont {Moon}}, \
  and\ \bibinfo {author} {\bibfnamefont {Y.-W.}\ \bibnamefont {Son}},\ }\href
  {\doibase 10.1103/PhysRevB.91.035405} {\bibfield  {journal} {\bibinfo
  {journal} {Phys. Rev. B}\ }\textbf {\bibinfo {volume} {91}},\ \bibinfo
  {pages} {035405} (\bibinfo {year} {2015})}\BibitemShut {NoStop}%
\bibitem [{\citenamefont {Zhao}\ \emph {et~al.}(2020)\citenamefont {Zhao},
  \citenamefont {Moon}, \citenamefont {Miyauchi}, \citenamefont {Nishihara},
  \citenamefont {Matsuda}, \citenamefont {Koshino},\ and\ \citenamefont
  {Kitaura}}]{PhysRevLett.124.106101}%
  \BibitemOpen
  \bibfield  {author} {\bibinfo {author} {\bibfnamefont {S.}~\bibnamefont
  {Zhao}}, \bibinfo {author} {\bibfnamefont {P.}~\bibnamefont {Moon}}, \bibinfo
  {author} {\bibfnamefont {Y.}~\bibnamefont {Miyauchi}}, \bibinfo {author}
  {\bibfnamefont {T.}~\bibnamefont {Nishihara}}, \bibinfo {author}
  {\bibfnamefont {K.}~\bibnamefont {Matsuda}}, \bibinfo {author} {\bibfnamefont
  {M.}~\bibnamefont {Koshino}}, \ and\ \bibinfo {author} {\bibfnamefont
  {R.}~\bibnamefont {Kitaura}},\ }\href {\doibase
  10.1103/PhysRevLett.124.106101} {\bibfield  {journal} {\bibinfo  {journal}
  {Phys. Rev. Lett.}\ }\textbf {\bibinfo {volume} {124}},\ \bibinfo {pages}
  {106101} (\bibinfo {year} {2020})}\BibitemShut {NoStop}%
\bibitem [{\citenamefont {Vu}\ and\ \citenamefont
  {Das~Sarma}(2021)}]{PhysRevLett.126.036803}%
  \BibitemOpen
  \bibfield  {author} {\bibinfo {author} {\bibfnamefont {D.}~\bibnamefont
  {Vu}}\ and\ \bibinfo {author} {\bibfnamefont {S.}~\bibnamefont {Das~Sarma}},\
  }\href {\doibase 10.1103/PhysRevLett.126.036803} {\bibfield  {journal}
  {\bibinfo  {journal} {Phys. Rev. Lett.}\ }\textbf {\bibinfo {volume} {126}},\
  \bibinfo {pages} {036803} (\bibinfo {year} {2021})}\BibitemShut {NoStop}%
\bibitem [{\citenamefont {Yu}\ \emph {et~al.}(2023)\citenamefont {Yu},
  \citenamefont {Li}, \citenamefont {Wang}, \citenamefont {Leykam},
  \citenamefont {Yuan},\ and\ \citenamefont {Chen}}]{PhysRevLett.130.143801}%
  \BibitemOpen
  \bibfield  {author} {\bibinfo {author} {\bibfnamefont {D.}~\bibnamefont
  {Yu}}, \bibinfo {author} {\bibfnamefont {G.}~\bibnamefont {Li}}, \bibinfo
  {author} {\bibfnamefont {L.}~\bibnamefont {Wang}}, \bibinfo {author}
  {\bibfnamefont {D.}~\bibnamefont {Leykam}}, \bibinfo {author} {\bibfnamefont
  {L.}~\bibnamefont {Yuan}}, \ and\ \bibinfo {author} {\bibfnamefont
  {X.}~\bibnamefont {Chen}},\ }\href {\doibase 10.1103/PhysRevLett.130.143801}
  {\bibfield  {journal} {\bibinfo  {journal} {Phys. Rev. Lett.}\ }\textbf
  {\bibinfo {volume} {130}},\ \bibinfo {pages} {143801} (\bibinfo {year}
  {2023})}\BibitemShut {NoStop}%
\bibitem [{\citenamefont {Nayak}\ \emph {et~al.}(2008)\citenamefont {Nayak},
  \citenamefont {Simon}, \citenamefont {Stern}, \citenamefont {Freedman},\ and\
  \citenamefont {Das~Sarma}}]{RevModPhys.80.1083}%
  \BibitemOpen
  \bibfield  {author} {\bibinfo {author} {\bibfnamefont {C.}~\bibnamefont
  {Nayak}}, \bibinfo {author} {\bibfnamefont {S.~H.}\ \bibnamefont {Simon}},
  \bibinfo {author} {\bibfnamefont {A.}~\bibnamefont {Stern}}, \bibinfo
  {author} {\bibfnamefont {M.}~\bibnamefont {Freedman}}, \ and\ \bibinfo
  {author} {\bibfnamefont {S.}~\bibnamefont {Das~Sarma}},\ }\href {\doibase
  10.1103/RevModPhys.80.1083} {\bibfield  {journal} {\bibinfo  {journal} {Rev.
  Mod. Phys.}\ }\textbf {\bibinfo {volume} {80}},\ \bibinfo {pages} {1083}
  (\bibinfo {year} {2008})}\BibitemShut {NoStop}%
\bibitem [{\citenamefont {He}\ \emph {et~al.}(2019)\citenamefont {He},
  \citenamefont {Sun},\ and\ \citenamefont {He}}]{he2019topological}%
  \BibitemOpen
  \bibfield  {author} {\bibinfo {author} {\bibfnamefont {M.}~\bibnamefont
  {He}}, \bibinfo {author} {\bibfnamefont {H.}~\bibnamefont {Sun}}, \ and\
  \bibinfo {author} {\bibfnamefont {Q.~L.}\ \bibnamefont {He}},\ }\href
  {\doibase 10.1007/s11467-019-0893-4} {\bibfield  {journal} {\bibinfo
  {journal} {Front. Phys.}\ }\textbf {\bibinfo {volume} {14}},\ \bibinfo
  {pages} {43401} (\bibinfo {year} {2019})}\BibitemShut {NoStop}%
\bibitem [{\citenamefont {Li}\ \emph {et~al.}(2024{\natexlab{a}})\citenamefont
  {Li}, \citenamefont {Xie}, \citenamefont {Li}, \citenamefont {Liang},
  \citenamefont {Li},\ and\ \citenamefont {Li}}]{Li2024scpma}%
  \BibitemOpen
  \bibfield  {author} {\bibinfo {author} {\bibfnamefont {J.}~\bibnamefont
  {Li}}, \bibinfo {author} {\bibfnamefont {Z.}~\bibnamefont {Xie}}, \bibinfo
  {author} {\bibfnamefont {Y.}~\bibnamefont {Li}}, \bibinfo {author}
  {\bibfnamefont {Y.}~\bibnamefont {Liang}}, \bibinfo {author} {\bibfnamefont
  {Z.}~\bibnamefont {Li}}, \ and\ \bibinfo {author} {\bibfnamefont
  {T.}~\bibnamefont {Li}},\ }\href {\doibase 10.1007/s11433-023-2245-9}
  {\bibfield  {journal} {\bibinfo  {journal} {Sci. China-Phys. Mech. Astron.}\
  }\textbf {\bibinfo {volume} {67}},\ \bibinfo {pages} {220311} (\bibinfo
  {year} {2024}{\natexlab{a}})}\BibitemShut {NoStop}%
\bibitem [{\citenamefont {Hasan}\ and\ \citenamefont
  {Kane}(2010)}]{RevModPhys.82.3045}%
  \BibitemOpen
  \bibfield  {author} {\bibinfo {author} {\bibfnamefont {M.~Z.}\ \bibnamefont
  {Hasan}}\ and\ \bibinfo {author} {\bibfnamefont {C.~L.}\ \bibnamefont
  {Kane}},\ }\href {\doibase 10.1103/RevModPhys.82.3045} {\bibfield  {journal}
  {\bibinfo  {journal} {Rev. Mod. Phys.}\ }\textbf {\bibinfo {volume} {82}},\
  \bibinfo {pages} {3045} (\bibinfo {year} {2010})}\BibitemShut {NoStop}%
\bibitem [{\citenamefont {Qi}\ and\ \citenamefont
  {Zhang}(2011)}]{RevModPhys.83.1057}%
  \BibitemOpen
  \bibfield  {author} {\bibinfo {author} {\bibfnamefont {X.-L.}\ \bibnamefont
  {Qi}}\ and\ \bibinfo {author} {\bibfnamefont {S.-C.}\ \bibnamefont {Zhang}},\
  }\href {\doibase 10.1103/RevModPhys.83.1057} {\bibfield  {journal} {\bibinfo
  {journal} {Rev. Mod. Phys.}\ }\textbf {\bibinfo {volume} {83}},\ \bibinfo
  {pages} {1057} (\bibinfo {year} {2011})}\BibitemShut {NoStop}%
\bibitem [{\citenamefont {Culcer}\ \emph {et~al.}(2020)\citenamefont {Culcer},
  \citenamefont {Keser}, \citenamefont {Li},\ and\ \citenamefont
  {Tkachov}}]{Culcer_2020}%
  \BibitemOpen
  \bibfield  {author} {\bibinfo {author} {\bibfnamefont {D.}~\bibnamefont
  {Culcer}}, \bibinfo {author} {\bibfnamefont {A.~C.}\ \bibnamefont {Keser}},
  \bibinfo {author} {\bibfnamefont {Y.}~\bibnamefont {Li}}, \ and\ \bibinfo
  {author} {\bibfnamefont {G.}~\bibnamefont {Tkachov}},\ }\href {\doibase
  10.1088/2053-1583/ab6ff7} {\bibfield  {journal} {\bibinfo  {journal} {2D
  Mater.}\ }\textbf {\bibinfo {volume} {7}},\ \bibinfo {pages} {022007}
  (\bibinfo {year} {2020})}\BibitemShut {NoStop}%
\bibitem [{\citenamefont {Armitage}\ \emph {et~al.}(2018)\citenamefont
  {Armitage}, \citenamefont {Mele},\ and\ \citenamefont
  {Vishwanath}}]{RevModPhys.90.015001}%
  \BibitemOpen
  \bibfield  {author} {\bibinfo {author} {\bibfnamefont {N.~P.}\ \bibnamefont
  {Armitage}}, \bibinfo {author} {\bibfnamefont {E.~J.}\ \bibnamefont {Mele}},
  \ and\ \bibinfo {author} {\bibfnamefont {A.}~\bibnamefont {Vishwanath}},\
  }\href {\doibase 10.1103/RevModPhys.90.015001} {\bibfield  {journal}
  {\bibinfo  {journal} {Rev. Mod. Phys.}\ }\textbf {\bibinfo {volume} {90}},\
  \bibinfo {pages} {015001} (\bibinfo {year} {2018})}\BibitemShut {NoStop}%
\bibitem [{\citenamefont {Lu}\ \emph {et~al.}(2014)\citenamefont {Lu},
  \citenamefont {Joannopoulos},\ and\ \citenamefont
  {Solja{\v{c}}i{\'c}}}]{lu2014topological}%
  \BibitemOpen
  \bibfield  {author} {\bibinfo {author} {\bibfnamefont {L.}~\bibnamefont
  {Lu}}, \bibinfo {author} {\bibfnamefont {J.~D.}\ \bibnamefont
  {Joannopoulos}}, \ and\ \bibinfo {author} {\bibfnamefont {M.}~\bibnamefont
  {Solja{\v{c}}i{\'c}}},\ }\href {\doibase 10.1038/nphoton.2014.248} {\bibfield
   {journal} {\bibinfo  {journal} {Nat. Photon.}\ }\textbf {\bibinfo {volume}
  {8}},\ \bibinfo {pages} {821} (\bibinfo {year} {2014})}\BibitemShut {NoStop}%
\bibitem [{\citenamefont {Roushan}\ \emph {et~al.}(2014)\citenamefont
  {Roushan}, \citenamefont {Neill}, \citenamefont {Chen}, \citenamefont
  {Kolodrubetz}, \citenamefont {Quintana}, \citenamefont {Leung}, \citenamefont
  {Fang}, \citenamefont {Barends}, \citenamefont {Campbell}, \citenamefont
  {Chen}, \citenamefont {Chiaro}, \citenamefont {Dunsworth}, \citenamefont
  {Jeffrey}, \citenamefont {Kelly}, \citenamefont {Megrant}, \citenamefont
  {Mutus}, \citenamefont {O'Malley}, \citenamefont {Sank}, \citenamefont
  {Vainsencher}, \citenamefont {Wenner}, \citenamefont {White}, \citenamefont
  {Polkovnikov}, \citenamefont {Cleland},\ and\ \citenamefont
  {Martinis}}]{roushan2014observation}%
  \BibitemOpen
  \bibfield  {author} {\bibinfo {author} {\bibfnamefont {P.}~\bibnamefont
  {Roushan}}, \bibinfo {author} {\bibfnamefont {C.}~\bibnamefont {Neill}},
  \bibinfo {author} {\bibfnamefont {Y.}~\bibnamefont {Chen}}, \bibinfo {author}
  {\bibfnamefont {M.}~\bibnamefont {Kolodrubetz}}, \bibinfo {author}
  {\bibfnamefont {C.}~\bibnamefont {Quintana}}, \bibinfo {author}
  {\bibfnamefont {N.}~\bibnamefont {Leung}}, \bibinfo {author} {\bibfnamefont
  {M.}~\bibnamefont {Fang}}, \bibinfo {author} {\bibfnamefont {R.}~\bibnamefont
  {Barends}}, \bibinfo {author} {\bibfnamefont {B.}~\bibnamefont {Campbell}},
  \bibinfo {author} {\bibfnamefont {Z.}~\bibnamefont {Chen}}, \bibinfo {author}
  {\bibfnamefont {B.}~\bibnamefont {Chiaro}}, \bibinfo {author} {\bibfnamefont
  {A.}~\bibnamefont {Dunsworth}}, \bibinfo {author} {\bibfnamefont
  {E.}~\bibnamefont {Jeffrey}}, \bibinfo {author} {\bibfnamefont
  {J.}~\bibnamefont {Kelly}}, \bibinfo {author} {\bibfnamefont
  {A.}~\bibnamefont {Megrant}}, \bibinfo {author} {\bibfnamefont
  {J.}~\bibnamefont {Mutus}}, \bibinfo {author} {\bibfnamefont {P.~J.~J.}\
  \bibnamefont {O'Malley}}, \bibinfo {author} {\bibfnamefont {D.}~\bibnamefont
  {Sank}}, \bibinfo {author} {\bibfnamefont {A.}~\bibnamefont {Vainsencher}},
  \bibinfo {author} {\bibfnamefont {J.}~\bibnamefont {Wenner}}, \bibinfo
  {author} {\bibfnamefont {T.}~\bibnamefont {White}}, \bibinfo {author}
  {\bibfnamefont {A.}~\bibnamefont {Polkovnikov}}, \bibinfo {author}
  {\bibfnamefont {A.~N.}\ \bibnamefont {Cleland}}, \ and\ \bibinfo {author}
  {\bibfnamefont {J.~M.}\ \bibnamefont {Martinis}},\ }\href {\doibase
  10.1038/nature13891} {\bibfield  {journal} {\bibinfo  {journal} {Nature}\
  }\textbf {\bibinfo {volume} {515}},\ \bibinfo {pages} {241} (\bibinfo {year}
  {2014})}\BibitemShut {NoStop}%
\bibitem [{\citenamefont {Schroer}\ \emph {et~al.}(2014)\citenamefont
  {Schroer}, \citenamefont {Kolodrubetz}, \citenamefont {Kindel}, \citenamefont
  {Sandberg}, \citenamefont {Gao}, \citenamefont {Vissers}, \citenamefont
  {Pappas}, \citenamefont {Polkovnikov},\ and\ \citenamefont
  {Lehnert}}]{PhysRevLett.113.050402}%
  \BibitemOpen
  \bibfield  {author} {\bibinfo {author} {\bibfnamefont {M.~D.}\ \bibnamefont
  {Schroer}}, \bibinfo {author} {\bibfnamefont {M.~H.}\ \bibnamefont
  {Kolodrubetz}}, \bibinfo {author} {\bibfnamefont {W.~F.}\ \bibnamefont
  {Kindel}}, \bibinfo {author} {\bibfnamefont {M.}~\bibnamefont {Sandberg}},
  \bibinfo {author} {\bibfnamefont {J.}~\bibnamefont {Gao}}, \bibinfo {author}
  {\bibfnamefont {M.~R.}\ \bibnamefont {Vissers}}, \bibinfo {author}
  {\bibfnamefont {D.~P.}\ \bibnamefont {Pappas}}, \bibinfo {author}
  {\bibfnamefont {A.}~\bibnamefont {Polkovnikov}}, \ and\ \bibinfo {author}
  {\bibfnamefont {K.~W.}\ \bibnamefont {Lehnert}},\ }\href {\doibase
  10.1103/PhysRevLett.113.050402} {\bibfield  {journal} {\bibinfo  {journal}
  {Phys. Rev. Lett.}\ }\textbf {\bibinfo {volume} {113}},\ \bibinfo {pages}
  {050402} (\bibinfo {year} {2014})}\BibitemShut {NoStop}%
\bibitem [{\citenamefont {Wei}\ \emph {et~al.}(2023)\citenamefont {Wei},
  \citenamefont {Wang}, \citenamefont {Wang}, \citenamefont {Xiang},
  \citenamefont {Xu}, \citenamefont {Wang},\ and\ \citenamefont
  {Wang}}]{PhysRevLett.130.036202}%
  \BibitemOpen
  \bibfield  {author} {\bibinfo {author} {\bibfnamefont {M.}~\bibnamefont
  {Wei}}, \bibinfo {author} {\bibfnamefont {L.}~\bibnamefont {Wang}}, \bibinfo
  {author} {\bibfnamefont {B.}~\bibnamefont {Wang}}, \bibinfo {author}
  {\bibfnamefont {L.}~\bibnamefont {Xiang}}, \bibinfo {author} {\bibfnamefont
  {F.}~\bibnamefont {Xu}}, \bibinfo {author} {\bibfnamefont {B.}~\bibnamefont
  {Wang}}, \ and\ \bibinfo {author} {\bibfnamefont {J.}~\bibnamefont {Wang}},\
  }\href {\doibase 10.1103/PhysRevLett.130.036202} {\bibfield  {journal}
  {\bibinfo  {journal} {Phys. Rev. Lett.}\ }\textbf {\bibinfo {volume} {130}},\
  \bibinfo {pages} {036202} (\bibinfo {year} {2023})}\BibitemShut {NoStop}%
\bibitem [{\citenamefont {Coen}\ \emph {et~al.}(2024)\citenamefont {Coen},
  \citenamefont {Garbin}, \citenamefont {Xu}, \citenamefont {Quinn},
  \citenamefont {Goldman}, \citenamefont {Oppo}, \citenamefont {Erkintalo},
  \citenamefont {Murdoch},\ and\ \citenamefont {Fatome}}]{coen2024nonlinear}%
  \BibitemOpen
  \bibfield  {author} {\bibinfo {author} {\bibfnamefont {S.}~\bibnamefont
  {Coen}}, \bibinfo {author} {\bibfnamefont {B.}~\bibnamefont {Garbin}},
  \bibinfo {author} {\bibfnamefont {G.}~\bibnamefont {Xu}}, \bibinfo {author}
  {\bibfnamefont {L.}~\bibnamefont {Quinn}}, \bibinfo {author} {\bibfnamefont
  {N.}~\bibnamefont {Goldman}}, \bibinfo {author} {\bibfnamefont {G.-L.}\
  \bibnamefont {Oppo}}, \bibinfo {author} {\bibfnamefont {M.}~\bibnamefont
  {Erkintalo}}, \bibinfo {author} {\bibfnamefont {S.~G.}\ \bibnamefont
  {Murdoch}}, \ and\ \bibinfo {author} {\bibfnamefont {J.}~\bibnamefont
  {Fatome}},\ }\href {\doibase 10.1038/s41467-023-44640-x} {\bibfield
  {journal} {\bibinfo  {journal} {Nat. Commun.}\ }\textbf {\bibinfo {volume}
  {15}},\ \bibinfo {pages} {1398} (\bibinfo {year} {2024})}\BibitemShut
  {NoStop}%
\bibitem [{\citenamefont {Tsui}\ \emph {et~al.}(1982)\citenamefont {Tsui},
  \citenamefont {Stormer},\ and\ \citenamefont
  {Gossard}}]{PhysRevLett.48.1559}%
  \BibitemOpen
  \bibfield  {author} {\bibinfo {author} {\bibfnamefont {D.~C.}\ \bibnamefont
  {Tsui}}, \bibinfo {author} {\bibfnamefont {H.~L.}\ \bibnamefont {Stormer}}, \
  and\ \bibinfo {author} {\bibfnamefont {A.~C.}\ \bibnamefont {Gossard}},\
  }\href {\doibase 10.1103/PhysRevLett.48.1559} {\bibfield  {journal} {\bibinfo
   {journal} {Phys. Rev. Lett.}\ }\textbf {\bibinfo {volume} {48}},\ \bibinfo
  {pages} {1559} (\bibinfo {year} {1982})}\BibitemShut {NoStop}%
\bibitem [{\citenamefont {Raghu}\ \emph {et~al.}(2008)\citenamefont {Raghu},
  \citenamefont {Qi}, \citenamefont {Honerkamp},\ and\ \citenamefont
  {Zhang}}]{PhysRevLett.100.156401}%
  \BibitemOpen
  \bibfield  {author} {\bibinfo {author} {\bibfnamefont {S.}~\bibnamefont
  {Raghu}}, \bibinfo {author} {\bibfnamefont {X.-L.}\ \bibnamefont {Qi}},
  \bibinfo {author} {\bibfnamefont {C.}~\bibnamefont {Honerkamp}}, \ and\
  \bibinfo {author} {\bibfnamefont {S.-C.}\ \bibnamefont {Zhang}},\ }\href
  {\doibase 10.1103/PhysRevLett.100.156401} {\bibfield  {journal} {\bibinfo
  {journal} {Phys. Rev. Lett.}\ }\textbf {\bibinfo {volume} {100}},\ \bibinfo
  {pages} {156401} (\bibinfo {year} {2008})}\BibitemShut {NoStop}%
\bibitem [{\citenamefont {Dauphin}\ \emph {et~al.}(2012)\citenamefont
  {Dauphin}, \citenamefont {M\"uller},\ and\ \citenamefont
  {Martin-Delgado}}]{PhysRevA.86.053618}%
  \BibitemOpen
  \bibfield  {author} {\bibinfo {author} {\bibfnamefont {A.}~\bibnamefont
  {Dauphin}}, \bibinfo {author} {\bibfnamefont {M.}~\bibnamefont {M\"uller}}, \
  and\ \bibinfo {author} {\bibfnamefont {M.~A.}\ \bibnamefont
  {Martin-Delgado}},\ }\href {\doibase 10.1103/PhysRevA.86.053618} {\bibfield
  {journal} {\bibinfo  {journal} {Phys. Rev. A}\ }\textbf {\bibinfo {volume}
  {86}},\ \bibinfo {pages} {053618} (\bibinfo {year} {2012})}\BibitemShut
  {NoStop}%
\bibitem [{\citenamefont {Xu}\ and\ \citenamefont
  {Chen}(2013)}]{PhysRevB.88.045110}%
  \BibitemOpen
  \bibfield  {author} {\bibinfo {author} {\bibfnamefont {Z.}~\bibnamefont
  {Xu}}\ and\ \bibinfo {author} {\bibfnamefont {S.}~\bibnamefont {Chen}},\
  }\href {\doibase 10.1103/PhysRevB.88.045110} {\bibfield  {journal} {\bibinfo
  {journal} {Phys. Rev. B}\ }\textbf {\bibinfo {volume} {88}},\ \bibinfo
  {pages} {045110} (\bibinfo {year} {2013})}\BibitemShut {NoStop}%
\bibitem [{\citenamefont {Zhang}\ \emph {et~al.}(2021)\citenamefont {Zhang},
  \citenamefont {Tang}, \citenamefont {Zhang}, \citenamefont {Zhang},\ and\
  \citenamefont {Zhu}}]{PhysRevB.104.L161118}%
  \BibitemOpen
  \bibfield  {author} {\bibinfo {author} {\bibfnamefont {G.-Q.}\ \bibnamefont
  {Zhang}}, \bibinfo {author} {\bibfnamefont {L.-Z.}\ \bibnamefont {Tang}},
  \bibinfo {author} {\bibfnamefont {L.-F.}\ \bibnamefont {Zhang}}, \bibinfo
  {author} {\bibfnamefont {D.-W.}\ \bibnamefont {Zhang}}, \ and\ \bibinfo
  {author} {\bibfnamefont {S.-L.}\ \bibnamefont {Zhu}},\ }\href {\doibase
  10.1103/PhysRevB.104.L161118} {\bibfield  {journal} {\bibinfo  {journal}
  {Phys. Rev. B}\ }\textbf {\bibinfo {volume} {104}},\ \bibinfo {pages}
  {L161118} (\bibinfo {year} {2021})}\BibitemShut {NoStop}%
\bibitem [{\citenamefont {Liu}\ \emph {et~al.}(2013)\citenamefont {Liu},
  \citenamefont {Liu},\ and\ \citenamefont {Cheng}}]{PhysRevLett.110.076401}%
  \BibitemOpen
  \bibfield  {author} {\bibinfo {author} {\bibfnamefont {X.-J.}\ \bibnamefont
  {Liu}}, \bibinfo {author} {\bibfnamefont {Z.-X.}\ \bibnamefont {Liu}}, \ and\
  \bibinfo {author} {\bibfnamefont {M.}~\bibnamefont {Cheng}},\ }\href
  {\doibase 10.1103/PhysRevLett.110.076401} {\bibfield  {journal} {\bibinfo
  {journal} {Phys. Rev. Lett.}\ }\textbf {\bibinfo {volume} {110}},\ \bibinfo
  {pages} {076401} (\bibinfo {year} {2013})}\BibitemShut {NoStop}%
\bibitem [{\citenamefont {Song}\ \emph {et~al.}(2018)\citenamefont {Song},
  \citenamefont {Zhang}, \citenamefont {He}, \citenamefont {Poon},
  \citenamefont {Hajiyev}, \citenamefont {Zhang}, \citenamefont {Liu},\ and\
  \citenamefont {Jo}}]{Song2018}%
  \BibitemOpen
  \bibfield  {author} {\bibinfo {author} {\bibfnamefont {B.}~\bibnamefont
  {Song}}, \bibinfo {author} {\bibfnamefont {L.}~\bibnamefont {Zhang}},
  \bibinfo {author} {\bibfnamefont {C.}~\bibnamefont {He}}, \bibinfo {author}
  {\bibfnamefont {T.~F.~J.}\ \bibnamefont {Poon}}, \bibinfo {author}
  {\bibfnamefont {E.}~\bibnamefont {Hajiyev}}, \bibinfo {author} {\bibfnamefont
  {S.}~\bibnamefont {Zhang}}, \bibinfo {author} {\bibfnamefont {X.-J.}\
  \bibnamefont {Liu}}, \ and\ \bibinfo {author} {\bibfnamefont {G.-B.}\
  \bibnamefont {Jo}},\ }\href {\doibase 10.1126/sciadv.aao4748} {\bibfield
  {journal} {\bibinfo  {journal} {Sci. Adv.}\ }\textbf {\bibinfo {volume}
  {4}},\ \bibinfo {pages} {eaao4748} (\bibinfo {year} {2018})}\BibitemShut
  {NoStop}%
\bibitem [{\citenamefont {Stuhl}\ \emph {et~al.}(2015)\citenamefont {Stuhl},
  \citenamefont {Lu}, \citenamefont {Aycock}, \citenamefont {Genkina},\ and\
  \citenamefont {Spielman}}]{doi:10.1126/science.aaa8515}%
  \BibitemOpen
  \bibfield  {author} {\bibinfo {author} {\bibfnamefont {B.~K.}\ \bibnamefont
  {Stuhl}}, \bibinfo {author} {\bibfnamefont {H.-I.}\ \bibnamefont {Lu}},
  \bibinfo {author} {\bibfnamefont {L.~M.}\ \bibnamefont {Aycock}}, \bibinfo
  {author} {\bibfnamefont {D.}~\bibnamefont {Genkina}}, \ and\ \bibinfo
  {author} {\bibfnamefont {I.~B.}\ \bibnamefont {Spielman}},\ }\href {\doibase
  10.1126/science.aaa8515} {\bibfield  {journal} {\bibinfo  {journal}
  {Science}\ }\textbf {\bibinfo {volume} {349}},\ \bibinfo {pages} {1514}
  (\bibinfo {year} {2015})}\BibitemShut {NoStop}%
\bibitem [{\citenamefont {Mancini}\ \emph {et~al.}(2015)\citenamefont
  {Mancini}, \citenamefont {Pagano}, \citenamefont {Cappellini}, \citenamefont
  {Livi}, \citenamefont {Rider}, \citenamefont {Catani}, \citenamefont {Sias},
  \citenamefont {Zoller}, \citenamefont {Inguscio}, \citenamefont {Dalmonte},\
  and\ \citenamefont {Fallani}}]{doi:10.1126/science.aaa8736}%
  \BibitemOpen
  \bibfield  {author} {\bibinfo {author} {\bibfnamefont {M.}~\bibnamefont
  {Mancini}}, \bibinfo {author} {\bibfnamefont {G.}~\bibnamefont {Pagano}},
  \bibinfo {author} {\bibfnamefont {G.}~\bibnamefont {Cappellini}}, \bibinfo
  {author} {\bibfnamefont {L.}~\bibnamefont {Livi}}, \bibinfo {author}
  {\bibfnamefont {M.}~\bibnamefont {Rider}}, \bibinfo {author} {\bibfnamefont
  {J.}~\bibnamefont {Catani}}, \bibinfo {author} {\bibfnamefont
  {C.}~\bibnamefont {Sias}}, \bibinfo {author} {\bibfnamefont {P.}~\bibnamefont
  {Zoller}}, \bibinfo {author} {\bibfnamefont {M.}~\bibnamefont {Inguscio}},
  \bibinfo {author} {\bibfnamefont {M.}~\bibnamefont {Dalmonte}}, \ and\
  \bibinfo {author} {\bibfnamefont {L.}~\bibnamefont {Fallani}},\ }\href
  {\doibase 10.1126/science.aaa8736} {\bibfield  {journal} {\bibinfo  {journal}
  {Science}\ }\textbf {\bibinfo {volume} {349}},\ \bibinfo {pages} {1510}
  (\bibinfo {year} {2015})}\BibitemShut {NoStop}%
\bibitem [{\citenamefont {Zhou}\ \emph {et~al.}(2023)\citenamefont {Zhou},
  \citenamefont {Wang}, \citenamefont {Poon}, \citenamefont {Zhou},\ and\
  \citenamefont {Liu}}]{PhysRevLett.131.176401}%
  \BibitemOpen
  \bibfield  {author} {\bibinfo {author} {\bibfnamefont {X.-C.}\ \bibnamefont
  {Zhou}}, \bibinfo {author} {\bibfnamefont {Y.}~\bibnamefont {Wang}}, \bibinfo
  {author} {\bibfnamefont {T.-F.~J.}\ \bibnamefont {Poon}}, \bibinfo {author}
  {\bibfnamefont {Q.}~\bibnamefont {Zhou}}, \ and\ \bibinfo {author}
  {\bibfnamefont {X.-J.}\ \bibnamefont {Liu}},\ }\href {\doibase
  10.1103/PhysRevLett.131.176401} {\bibfield  {journal} {\bibinfo  {journal}
  {Phys. Rev. Lett.}\ }\textbf {\bibinfo {volume} {131}},\ \bibinfo {pages}
  {176401} (\bibinfo {year} {2023})}\BibitemShut {NoStop}%
\bibitem [{\citenamefont {Tang}\ \emph {et~al.}(2022)\citenamefont {Tang},
  \citenamefont {Liu}, \citenamefont {Zhang},\ and\ \citenamefont
  {Zhang}}]{PhysRevA.105.063327}%
  \BibitemOpen
  \bibfield  {author} {\bibinfo {author} {\bibfnamefont {L.-Z.}\ \bibnamefont
  {Tang}}, \bibinfo {author} {\bibfnamefont {S.-N.}\ \bibnamefont {Liu}},
  \bibinfo {author} {\bibfnamefont {G.-Q.}\ \bibnamefont {Zhang}}, \ and\
  \bibinfo {author} {\bibfnamefont {D.-W.}\ \bibnamefont {Zhang}},\ }\href
  {\doibase 10.1103/PhysRevA.105.063327} {\bibfield  {journal} {\bibinfo
  {journal} {Phys. Rev. A}\ }\textbf {\bibinfo {volume} {105}},\ \bibinfo
  {pages} {063327} (\bibinfo {year} {2022})}\BibitemShut {NoStop}%
\bibitem [{\citenamefont {Li}\ \emph {et~al.}(2024{\natexlab{b}})\citenamefont
  {Li}, \citenamefont {Xu}, \citenamefont {Wang}, \citenamefont {Tang},
  \citenamefont {Zhang}, \citenamefont {Yang}, \citenamefont {Su},
  \citenamefont {Wang}, \citenamefont {Mi}, \citenamefont {Sun}, \citenamefont
  {Liang}, \citenamefont {Chen}, \citenamefont {Li}, \citenamefont {Zhang},
  \citenamefont {Linghu}, \citenamefont {Han}, \citenamefont {Liu},
  \citenamefont {Feng}, \citenamefont {Liu}, \citenamefont {Xue}, \citenamefont
  {Zhang}, \citenamefont {Jin}, \citenamefont {Zhu}, \citenamefont {Yu},
  \citenamefont {Zhao},\ and\ \citenamefont {Xue}}]{Li2024}%
  \BibitemOpen
  \bibfield  {author} {\bibinfo {author} {\bibfnamefont {X.}~\bibnamefont
  {Li}}, \bibinfo {author} {\bibfnamefont {H.}~\bibnamefont {Xu}}, \bibinfo
  {author} {\bibfnamefont {J.}~\bibnamefont {Wang}}, \bibinfo {author}
  {\bibfnamefont {L.-Z.}\ \bibnamefont {Tang}}, \bibinfo {author}
  {\bibfnamefont {D.-W.}\ \bibnamefont {Zhang}}, \bibinfo {author}
  {\bibfnamefont {C.}~\bibnamefont {Yang}}, \bibinfo {author} {\bibfnamefont
  {T.}~\bibnamefont {Su}}, \bibinfo {author} {\bibfnamefont {C.}~\bibnamefont
  {Wang}}, \bibinfo {author} {\bibfnamefont {Z.}~\bibnamefont {Mi}}, \bibinfo
  {author} {\bibfnamefont {W.}~\bibnamefont {Sun}}, \bibinfo {author}
  {\bibfnamefont {X.}~\bibnamefont {Liang}}, \bibinfo {author} {\bibfnamefont
  {M.}~\bibnamefont {Chen}}, \bibinfo {author} {\bibfnamefont {C.}~\bibnamefont
  {Li}}, \bibinfo {author} {\bibfnamefont {Y.}~\bibnamefont {Zhang}}, \bibinfo
  {author} {\bibfnamefont {K.}~\bibnamefont {Linghu}}, \bibinfo {author}
  {\bibfnamefont {J.}~\bibnamefont {Han}}, \bibinfo {author} {\bibfnamefont
  {W.}~\bibnamefont {Liu}}, \bibinfo {author} {\bibfnamefont {Y.}~\bibnamefont
  {Feng}}, \bibinfo {author} {\bibfnamefont {P.}~\bibnamefont {Liu}}, \bibinfo
  {author} {\bibfnamefont {G.}~\bibnamefont {Xue}}, \bibinfo {author}
  {\bibfnamefont {J.}~\bibnamefont {Zhang}}, \bibinfo {author} {\bibfnamefont
  {Y.}~\bibnamefont {Jin}}, \bibinfo {author} {\bibfnamefont {S.-L.}\
  \bibnamefont {Zhu}}, \bibinfo {author} {\bibfnamefont {H.}~\bibnamefont
  {Yu}}, \bibinfo {author} {\bibfnamefont {S.~P.}\ \bibnamefont {Zhao}}, \ and\
  \bibinfo {author} {\bibfnamefont {Q.-K.}\ \bibnamefont {Xue}},\ }\href
  {\doibase 10.1103/physrevresearch.6.l042038} {\bibfield  {journal} {\bibinfo
  {journal} {Phys. Rev. Res.}\ }\textbf {\bibinfo {volume} {6}},\ \bibinfo
  {pages} {1042038} (\bibinfo {year} {2024}{\natexlab{b}})}\BibitemShut
  {NoStop}%
\bibitem [{\citenamefont {Li}\ \emph {et~al.}(2009)\citenamefont {Li},
  \citenamefont {Chu}, \citenamefont {Jain},\ and\ \citenamefont
  {Shen}}]{PhysRevLett.102.136806}%
  \BibitemOpen
  \bibfield  {author} {\bibinfo {author} {\bibfnamefont {J.}~\bibnamefont
  {Li}}, \bibinfo {author} {\bibfnamefont {R.-L.}\ \bibnamefont {Chu}},
  \bibinfo {author} {\bibfnamefont {J.~K.}\ \bibnamefont {Jain}}, \ and\
  \bibinfo {author} {\bibfnamefont {S.-Q.}\ \bibnamefont {Shen}},\ }\href
  {\doibase 10.1103/PhysRevLett.102.136806} {\bibfield  {journal} {\bibinfo
  {journal} {Phys. Rev. Lett.}\ }\textbf {\bibinfo {volume} {102}},\ \bibinfo
  {pages} {136806} (\bibinfo {year} {2009})}\BibitemShut {NoStop}%
\bibitem [{\citenamefont {Groth}\ \emph {et~al.}(2009)\citenamefont {Groth},
  \citenamefont {Wimmer}, \citenamefont {Akhmerov}, \citenamefont
  {Tworzyd\l{}o},\ and\ \citenamefont {Beenakker}}]{PhysRevLett.103.196805}%
  \BibitemOpen
  \bibfield  {author} {\bibinfo {author} {\bibfnamefont {C.~W.}\ \bibnamefont
  {Groth}}, \bibinfo {author} {\bibfnamefont {M.}~\bibnamefont {Wimmer}},
  \bibinfo {author} {\bibfnamefont {A.~R.}\ \bibnamefont {Akhmerov}}, \bibinfo
  {author} {\bibfnamefont {J.}~\bibnamefont {Tworzyd\l{}o}}, \ and\ \bibinfo
  {author} {\bibfnamefont {C.~W.~J.}\ \bibnamefont {Beenakker}},\ }\href
  {\doibase 10.1103/PhysRevLett.103.196805} {\bibfield  {journal} {\bibinfo
  {journal} {Phys. Rev. Lett.}\ }\textbf {\bibinfo {volume} {103}},\ \bibinfo
  {pages} {196805} (\bibinfo {year} {2009})}\BibitemShut {NoStop}%
\bibitem [{\citenamefont {Zhang}\ \emph
  {et~al.}(2020{\natexlab{b}})\citenamefont {Zhang}, \citenamefont {Tang},
  \citenamefont {Lang}, \citenamefont {Yan},\ and\ \citenamefont
  {Zhu}}]{Zhang2020}%
  \BibitemOpen
  \bibfield  {author} {\bibinfo {author} {\bibfnamefont {D.-W.}\ \bibnamefont
  {Zhang}}, \bibinfo {author} {\bibfnamefont {L.-Z.}\ \bibnamefont {Tang}},
  \bibinfo {author} {\bibfnamefont {L.-J.}\ \bibnamefont {Lang}}, \bibinfo
  {author} {\bibfnamefont {H.}~\bibnamefont {Yan}}, \ and\ \bibinfo {author}
  {\bibfnamefont {S.-L.}\ \bibnamefont {Zhu}},\ }\href {\doibase
  10.1007/s11433-020-1521-9} {\bibfield  {journal} {\bibinfo  {journal} {Sci.
  China-Phys. Mech. Astron.}\ }\textbf {\bibinfo {volume} {63}},\ \bibinfo
  {pages} {267062} (\bibinfo {year} {2020}{\natexlab{b}})}\BibitemShut
  {NoStop}%
\bibitem [{\citenamefont {Gu}\ \emph {et~al.}(2023)\citenamefont {Gu},
  \citenamefont {Gao}, \citenamefont {Xue}, \citenamefont {Wang}, \citenamefont
  {Guo}, \citenamefont {Su}, \citenamefont {Zhang},\ and\ \citenamefont
  {Zhu}}]{Gu2023}%
  \BibitemOpen
  \bibfield  {author} {\bibinfo {author} {\bibfnamefont {Z.}~\bibnamefont
  {Gu}}, \bibinfo {author} {\bibfnamefont {H.}~\bibnamefont {Gao}}, \bibinfo
  {author} {\bibfnamefont {H.}~\bibnamefont {Xue}}, \bibinfo {author}
  {\bibfnamefont {D.}~\bibnamefont {Wang}}, \bibinfo {author} {\bibfnamefont
  {J.}~\bibnamefont {Guo}}, \bibinfo {author} {\bibfnamefont {Z.}~\bibnamefont
  {Su}}, \bibinfo {author} {\bibfnamefont {B.}~\bibnamefont {Zhang}}, \ and\
  \bibinfo {author} {\bibfnamefont {J.}~\bibnamefont {Zhu}},\ }\href {\doibase
  10.1007/s11433-023-2159-4} {\bibfield  {journal} {\bibinfo  {journal} {Sci.
  China-Phys. Mech. Astron.}\ }\textbf {\bibinfo {volume} {66}},\ \bibinfo
  {pages} {294311} (\bibinfo {year} {2023})}\BibitemShut {NoStop}%
\bibitem [{\citenamefont {Chiu}\ \emph {et~al.}(2016)\citenamefont {Chiu},
  \citenamefont {Teo}, \citenamefont {Schnyder},\ and\ \citenamefont
  {Ryu}}]{RevModPhys.88.035005}%
  \BibitemOpen
  \bibfield  {author} {\bibinfo {author} {\bibfnamefont {C.-K.}\ \bibnamefont
  {Chiu}}, \bibinfo {author} {\bibfnamefont {J.~C.~Y.}\ \bibnamefont {Teo}},
  \bibinfo {author} {\bibfnamefont {A.~P.}\ \bibnamefont {Schnyder}}, \ and\
  \bibinfo {author} {\bibfnamefont {S.}~\bibnamefont {Ryu}},\ }\href {\doibase
  10.1103/RevModPhys.88.035005} {\bibfield  {journal} {\bibinfo  {journal}
  {Rev. Mod. Phys.}\ }\textbf {\bibinfo {volume} {88}},\ \bibinfo {pages}
  {035005} (\bibinfo {year} {2016})}\BibitemShut {NoStop}%
\bibitem [{\citenamefont {Mondragon-Shem}\ \emph {et~al.}(2014)\citenamefont
  {Mondragon-Shem}, \citenamefont {Hughes}, \citenamefont {Song},\ and\
  \citenamefont {Prodan}}]{PhysRevLett.113.046802}%
  \BibitemOpen
  \bibfield  {author} {\bibinfo {author} {\bibfnamefont {I.}~\bibnamefont
  {Mondragon-Shem}}, \bibinfo {author} {\bibfnamefont {T.~L.}\ \bibnamefont
  {Hughes}}, \bibinfo {author} {\bibfnamefont {J.}~\bibnamefont {Song}}, \ and\
  \bibinfo {author} {\bibfnamefont {E.}~\bibnamefont {Prodan}},\ }\href
  {\doibase 10.1103/PhysRevLett.113.046802} {\bibfield  {journal} {\bibinfo
  {journal} {Phys. Rev. Lett.}\ }\textbf {\bibinfo {volume} {113}},\ \bibinfo
  {pages} {046802} (\bibinfo {year} {2014})}\BibitemShut {NoStop}%
\bibitem [{\citenamefont {Peschel}\ and\ \citenamefont
  {Eisler}(2009)}]{Peschel_2009}%
  \BibitemOpen
  \bibfield  {author} {\bibinfo {author} {\bibfnamefont {I.}~\bibnamefont
  {Peschel}}\ and\ \bibinfo {author} {\bibfnamefont {V.}~\bibnamefont
  {Eisler}},\ }\href {\doibase 10.1088/1751-8113/42/50/504003} {\bibfield
  {journal} {\bibinfo  {journal} {J. Phys. A: Math. Theor.}\ }\textbf {\bibinfo
  {volume} {42}},\ \bibinfo {pages} {504003} (\bibinfo {year}
  {2009})}\BibitemShut {NoStop}%
\bibitem [{\citenamefont {Zhou}(2022)}]{PhysRevResearch.4.043164}%
  \BibitemOpen
  \bibfield  {author} {\bibinfo {author} {\bibfnamefont {L.}~\bibnamefont
  {Zhou}},\ }\href {\doibase 10.1103/PhysRevResearch.4.043164} {\bibfield
  {journal} {\bibinfo  {journal} {Phys. Rev. Res.}\ }\textbf {\bibinfo {volume}
  {4}},\ \bibinfo {pages} {043164} (\bibinfo {year} {2022})}\BibitemShut
  {NoStop}%
\bibitem [{\citenamefont {Yu}\ \emph {et~al.}(2024)\citenamefont {Yu},
  \citenamefont {Yang}, \citenamefont {Lin},\ and\ \citenamefont
  {Jian}}]{PhysRevLett.133.026601}%
  \BibitemOpen
  \bibfield  {author} {\bibinfo {author} {\bibfnamefont {X.-J.}\ \bibnamefont
  {Yu}}, \bibinfo {author} {\bibfnamefont {S.}~\bibnamefont {Yang}}, \bibinfo
  {author} {\bibfnamefont {H.-Q.}\ \bibnamefont {Lin}}, \ and\ \bibinfo
  {author} {\bibfnamefont {S.-K.}\ \bibnamefont {Jian}},\ }\href {\doibase
  10.1103/PhysRevLett.133.026601} {\bibfield  {journal} {\bibinfo  {journal}
  {Phys. Rev. Lett.}\ }\textbf {\bibinfo {volume} {133}},\ \bibinfo {pages}
  {026601} (\bibinfo {year} {2024})}\BibitemShut {NoStop}%
\bibitem [{\citenamefont {Kumar}\ \emph {et~al.}(2025)\citenamefont {Kumar},
  \citenamefont {Vedula}, \citenamefont {Gangadharaiah},\ and\ \citenamefont
  {Sharma}}]{kumar2025entanglemententropyprobetopological}%
  \BibitemOpen
  \bibfield  {author} {\bibinfo {author} {\bibfnamefont {M.}~\bibnamefont
  {Kumar}}, \bibinfo {author} {\bibfnamefont {B.}~\bibnamefont {Vedula}},
  \bibinfo {author} {\bibfnamefont {S.}~\bibnamefont {Gangadharaiah}}, \ and\
  \bibinfo {author} {\bibfnamefont {A.}~\bibnamefont {Sharma}},\ }\href
  {https://arxiv.org/abs/2508.15897} {\bibfield  {journal} {\bibinfo  {journal}
  {arXiv preprint arXiv:2508.15897}\ } (\bibinfo {year} {2025})}\BibitemShut
  {NoStop}%
\bibitem [{\citenamefont {Yin}\ \emph {et~al.}(2018)\citenamefont {Yin},
  \citenamefont {Jiang}, \citenamefont {Li}, \citenamefont {L\"u},\ and\
  \citenamefont {Chen}}]{PhysRevA.97.052115}%
  \BibitemOpen
  \bibfield  {author} {\bibinfo {author} {\bibfnamefont {C.}~\bibnamefont
  {Yin}}, \bibinfo {author} {\bibfnamefont {H.}~\bibnamefont {Jiang}}, \bibinfo
  {author} {\bibfnamefont {L.}~\bibnamefont {Li}}, \bibinfo {author}
  {\bibfnamefont {R.}~\bibnamefont {L\"u}}, \ and\ \bibinfo {author}
  {\bibfnamefont {S.}~\bibnamefont {Chen}},\ }\href {\doibase
  10.1103/PhysRevA.97.052115} {\bibfield  {journal} {\bibinfo  {journal} {Phys.
  Rev. A}\ }\textbf {\bibinfo {volume} {97}},\ \bibinfo {pages} {052115}
  (\bibinfo {year} {2018})}\BibitemShut {NoStop}%
\bibitem [{\citenamefont {Osterloh}\ \emph {et~al.}(2002)\citenamefont
  {Osterloh}, \citenamefont {Amico}, \citenamefont {Falci},\ and\ \citenamefont
  {Fazio}}]{Osterloh2002}%
  \BibitemOpen
  \bibfield  {author} {\bibinfo {author} {\bibfnamefont {A.}~\bibnamefont
  {Osterloh}}, \bibinfo {author} {\bibfnamefont {L.}~\bibnamefont {Amico}},
  \bibinfo {author} {\bibfnamefont {G.}~\bibnamefont {Falci}}, \ and\ \bibinfo
  {author} {\bibfnamefont {R.}~\bibnamefont {Fazio}},\ }\href {\doibase
  10.1038/416608a} {\bibfield  {journal} {\bibinfo  {journal} {Nature}\
  }\textbf {\bibinfo {volume} {416}},\ \bibinfo {pages} {608} (\bibinfo {year}
  {2002})}\BibitemShut {NoStop}%
\bibitem [{\citenamefont {Gu}\ \emph {et~al.}(2008)\citenamefont {Gu},
  \citenamefont {Kwok}, \citenamefont {Ning},\ and\ \citenamefont
  {Lin}}]{PhysRevB.77.245109}%
  \BibitemOpen
  \bibfield  {author} {\bibinfo {author} {\bibfnamefont {S.-J.}\ \bibnamefont
  {Gu}}, \bibinfo {author} {\bibfnamefont {H.-M.}\ \bibnamefont {Kwok}},
  \bibinfo {author} {\bibfnamefont {W.-Q.}\ \bibnamefont {Ning}}, \ and\
  \bibinfo {author} {\bibfnamefont {H.-Q.}\ \bibnamefont {Lin}},\ }\href
  {\doibase 10.1103/PhysRevB.77.245109} {\bibfield  {journal} {\bibinfo
  {journal} {Phys. Rev. B}\ }\textbf {\bibinfo {volume} {77}},\ \bibinfo
  {pages} {245109} (\bibinfo {year} {2008})}\BibitemShut {NoStop}%
\bibitem [{\citenamefont {Albuquerque}\ \emph {et~al.}(2010)\citenamefont
  {Albuquerque}, \citenamefont {Alet}, \citenamefont {Sire},\ and\
  \citenamefont {Capponi}}]{PhysRevB.81.064418}%
  \BibitemOpen
  \bibfield  {author} {\bibinfo {author} {\bibfnamefont {A.~F.}\ \bibnamefont
  {Albuquerque}}, \bibinfo {author} {\bibfnamefont {F.}~\bibnamefont {Alet}},
  \bibinfo {author} {\bibfnamefont {C.}~\bibnamefont {Sire}}, \ and\ \bibinfo
  {author} {\bibfnamefont {S.}~\bibnamefont {Capponi}},\ }\href {\doibase
  10.1103/PhysRevB.81.064418} {\bibfield  {journal} {\bibinfo  {journal} {Phys.
  Rev. B}\ }\textbf {\bibinfo {volume} {81}},\ \bibinfo {pages} {064418}
  (\bibinfo {year} {2010})}\BibitemShut {NoStop}%
\bibitem [{\citenamefont {Chen}(2016)}]{Chen2016}%
  \BibitemOpen
  \bibfield  {author} {\bibinfo {author} {\bibfnamefont {W.}~\bibnamefont
  {Chen}},\ }\href {\doibase 10.1088/0953-8984/28/5/055601} {\bibfield
  {journal} {\bibinfo  {journal} {J. Phys. Condens. Matter}\ }\textbf {\bibinfo
  {volume} {28}},\ \bibinfo {pages} {055601} (\bibinfo {year}
  {2016})}\BibitemShut {NoStop}%
\bibitem [{\citenamefont {Chen}\ \emph {et~al.}(2017)\citenamefont {Chen},
  \citenamefont {Legner}, \citenamefont {R\"uegg},\ and\ \citenamefont
  {Sigrist}}]{PhysRevB.95.075116}%
  \BibitemOpen
  \bibfield  {author} {\bibinfo {author} {\bibfnamefont {W.}~\bibnamefont
  {Chen}}, \bibinfo {author} {\bibfnamefont {M.}~\bibnamefont {Legner}},
  \bibinfo {author} {\bibfnamefont {A.}~\bibnamefont {R\"uegg}}, \ and\
  \bibinfo {author} {\bibfnamefont {M.}~\bibnamefont {Sigrist}},\ }\href
  {\doibase 10.1103/PhysRevB.95.075116} {\bibfield  {journal} {\bibinfo
  {journal} {Phys. Rev. B}\ }\textbf {\bibinfo {volume} {95}},\ \bibinfo
  {pages} {075116} (\bibinfo {year} {2017})}\BibitemShut {NoStop}%
\bibitem [{\citenamefont {Yu}\ and\ \citenamefont
  {Li}(2024)}]{PhysRevB.110.045119}%
  \BibitemOpen
  \bibfield  {author} {\bibinfo {author} {\bibfnamefont {X.-J.}\ \bibnamefont
  {Yu}}\ and\ \bibinfo {author} {\bibfnamefont {W.-L.}\ \bibnamefont {Li}},\
  }\href {\doibase 10.1103/PhysRevB.110.045119} {\bibfield  {journal} {\bibinfo
   {journal} {Phys. Rev. B}\ }\textbf {\bibinfo {volume} {110}},\ \bibinfo
  {pages} {045119} (\bibinfo {year} {2024})}\BibitemShut {NoStop}%
\bibitem [{\citenamefont {Liang}\ \emph {et~al.}(2024)\citenamefont {Liang},
  \citenamefont {Tang},\ and\ \citenamefont {Zhang}}]{PhysRevB.110.024207}%
  \BibitemOpen
  \bibfield  {author} {\bibinfo {author} {\bibfnamefont {E.-W.}\ \bibnamefont
  {Liang}}, \bibinfo {author} {\bibfnamefont {L.-Z.}\ \bibnamefont {Tang}}, \
  and\ \bibinfo {author} {\bibfnamefont {D.-W.}\ \bibnamefont {Zhang}},\ }\href
  {\doibase 10.1103/PhysRevB.110.024207} {\bibfield  {journal} {\bibinfo
  {journal} {Phys. Rev. B}\ }\textbf {\bibinfo {volume} {110}},\ \bibinfo
  {pages} {024207} (\bibinfo {year} {2024})}\BibitemShut {NoStop}%
\bibitem [{\citenamefont {Assun{\c{c}}{\~a}o}\ \emph
  {et~al.}(2024)\citenamefont {Assun{\c{c}}{\~a}o}, \citenamefont {Ferreira},\
  and\ \citenamefont {Lewenkopf}}]{PhysRevB.109.L201102}%
  \BibitemOpen
  \bibfield  {author} {\bibinfo {author} {\bibfnamefont {B.~D.}\ \bibnamefont
  {Assun{\c{c}}{\~a}o}}, \bibinfo {author} {\bibfnamefont {G.~J.}\ \bibnamefont
  {Ferreira}}, \ and\ \bibinfo {author} {\bibfnamefont {C.~H.}\ \bibnamefont
  {Lewenkopf}},\ }\href {\doibase 10.1103/PhysRevB.109.L201102} {\bibfield
  {journal} {\bibinfo  {journal} {Phys. Rev. B}\ }\textbf {\bibinfo {volume}
  {109}},\ \bibinfo {pages} {L201102} (\bibinfo {year} {2024})}\BibitemShut
  {NoStop}%
\bibitem [{\citenamefont {Calabrese}\ and\ \citenamefont
  {Cardy}(2004)}]{PasqualeCalabrese_2004}%
  \BibitemOpen
  \bibfield  {author} {\bibinfo {author} {\bibfnamefont {P.}~\bibnamefont
  {Calabrese}}\ and\ \bibinfo {author} {\bibfnamefont {J.}~\bibnamefont
  {Cardy}},\ }\href {\doibase 10.1088/1742-5468/2004/06/P06002} {\bibfield
  {journal} {\bibinfo  {journal} {J. Stat. Mech.: Theory Exp.}\ }\textbf
  {\bibinfo {volume} {2004}},\ \bibinfo {pages} {P06002} (\bibinfo {year}
  {2004})}\BibitemShut {NoStop}%
\bibitem [{\citenamefont {Calabrese}\ and\ \citenamefont
  {Cardy}(2009)}]{Calabrese_2009}%
  \BibitemOpen
  \bibfield  {author} {\bibinfo {author} {\bibfnamefont {P.}~\bibnamefont
  {Calabrese}}\ and\ \bibinfo {author} {\bibfnamefont {J.}~\bibnamefont
  {Cardy}},\ }\href {\doibase 10.1088/1751-8113/42/50/504005} {\bibfield
  {journal} {\bibinfo  {journal} {J. Phys. A}\ }\textbf {\bibinfo {volume}
  {42}},\ \bibinfo {pages} {504005} (\bibinfo {year} {2009})}\BibitemShut
  {NoStop}%
\bibitem [{\citenamefont {Voit}(1995)}]{JVoit_1995}%
  \BibitemOpen
  \bibfield  {author} {\bibinfo {author} {\bibfnamefont {J.}~\bibnamefont
  {Voit}},\ }\href {\doibase 10.1088/0034-4885/58/9/002} {\bibfield  {journal}
  {\bibinfo  {journal} {Rep. Prog. Phys.}\ }\textbf {\bibinfo {volume} {58}},\
  \bibinfo {pages} {977} (\bibinfo {year} {1995})}\BibitemShut {NoStop}%
\bibitem [{\citenamefont {Xie}\ \emph {et~al.}(2019)\citenamefont {Xie},
  \citenamefont {Gou}, \citenamefont {Xiao}, \citenamefont {Gadway},\ and\
  \citenamefont {Yan}}]{Xie2019}%
  \BibitemOpen
  \bibfield  {author} {\bibinfo {author} {\bibfnamefont {D.}~\bibnamefont
  {Xie}}, \bibinfo {author} {\bibfnamefont {W.}~\bibnamefont {Gou}}, \bibinfo
  {author} {\bibfnamefont {T.}~\bibnamefont {Xiao}}, \bibinfo {author}
  {\bibfnamefont {B.}~\bibnamefont {Gadway}}, \ and\ \bibinfo {author}
  {\bibfnamefont {B.}~\bibnamefont {Yan}},\ }\href {\doibase
  10.1038/s41534-019-0159-6} {\bibfield  {journal} {\bibinfo  {journal} {npj
  Quantum Inf.}\ }\textbf {\bibinfo {volume} {5}},\ \bibinfo {pages} {55}
  (\bibinfo {year} {2019})}\BibitemShut {NoStop}%
\bibitem [{\citenamefont {Gadway}(2015)}]{PhysRevA.92.043606}%
  \BibitemOpen
  \bibfield  {author} {\bibinfo {author} {\bibfnamefont {B.}~\bibnamefont
  {Gadway}},\ }\href {\doibase 10.1103/PhysRevA.92.043606} {\bibfield
  {journal} {\bibinfo  {journal} {Phys. Rev. A}\ }\textbf {\bibinfo {volume}
  {92}},\ \bibinfo {pages} {043606} (\bibinfo {year} {2015})}\BibitemShut
  {NoStop}%
\bibitem [{\citenamefont {Meier}\ \emph {et~al.}(2018)\citenamefont {Meier},
  \citenamefont {An}, \citenamefont {Dauphin}, \citenamefont {Maffei},
  \citenamefont {Massignan}, \citenamefont {Hughes},\ and\ \citenamefont
  {Gadway}}]{doi:10.1126/science.aat3406}%
  \BibitemOpen
  \bibfield  {author} {\bibinfo {author} {\bibfnamefont {E.~J.}\ \bibnamefont
  {Meier}}, \bibinfo {author} {\bibfnamefont {F.~A.}\ \bibnamefont {An}},
  \bibinfo {author} {\bibfnamefont {A.}~\bibnamefont {Dauphin}}, \bibinfo
  {author} {\bibfnamefont {M.}~\bibnamefont {Maffei}}, \bibinfo {author}
  {\bibfnamefont {P.}~\bibnamefont {Massignan}}, \bibinfo {author}
  {\bibfnamefont {T.~L.}\ \bibnamefont {Hughes}}, \ and\ \bibinfo {author}
  {\bibfnamefont {B.}~\bibnamefont {Gadway}},\ }\href {\doibase
  10.1126/science.aat3406} {\bibfield  {journal} {\bibinfo  {journal}
  {Science}\ }\textbf {\bibinfo {volume} {362}},\ \bibinfo {pages} {929}
  (\bibinfo {year} {2018})}\BibitemShut {NoStop}%
\bibitem [{\citenamefont {Li}\ \emph {et~al.}(2023)\citenamefont {Li},
  \citenamefont {Du}, \citenamefont {Wang}, \citenamefont {Liang},
  \citenamefont {Xiao}, \citenamefont {Yi}, \citenamefont {Ma},\ and\
  \citenamefont {Jia}}]{Li2023}%
  \BibitemOpen
  \bibfield  {author} {\bibinfo {author} {\bibfnamefont {Y.}~\bibnamefont
  {Li}}, \bibinfo {author} {\bibfnamefont {H.}~\bibnamefont {Du}}, \bibinfo
  {author} {\bibfnamefont {Y.}~\bibnamefont {Wang}}, \bibinfo {author}
  {\bibfnamefont {J.}~\bibnamefont {Liang}}, \bibinfo {author} {\bibfnamefont
  {L.}~\bibnamefont {Xiao}}, \bibinfo {author} {\bibfnamefont {W.}~\bibnamefont
  {Yi}}, \bibinfo {author} {\bibfnamefont {J.}~\bibnamefont {Ma}}, \ and\
  \bibinfo {author} {\bibfnamefont {S.}~\bibnamefont {Jia}},\ }\href {\doibase
  10.1038/s41467-023-43204-3} {\bibfield  {journal} {\bibinfo  {journal} {Nat.
  Commun.}\ }\textbf {\bibinfo {volume} {14}},\ \bibinfo {pages} {7560}
  (\bibinfo {year} {2023})}\BibitemShut {NoStop}%
\bibitem [{\citenamefont {Ren}\ \emph {et~al.}(2023)\citenamefont {Ren},
  \citenamefont {Li}, \citenamefont {Wu}, \citenamefont {Zhao}, \citenamefont
  {Wang}, \citenamefont {Liu}, \citenamefont {Li}, \citenamefont {Fu},
  \citenamefont {Xiao}, \citenamefont {Ma},\ and\ \citenamefont
  {Jia}}]{Ren:23}%
  \BibitemOpen
  \bibfield  {author} {\bibinfo {author} {\bibfnamefont {C.}~\bibnamefont
  {Ren}}, \bibinfo {author} {\bibfnamefont {Y.}~\bibnamefont {Li}}, \bibinfo
  {author} {\bibfnamefont {J.}~\bibnamefont {Wu}}, \bibinfo {author}
  {\bibfnamefont {H.}~\bibnamefont {Zhao}}, \bibinfo {author} {\bibfnamefont
  {Y.}~\bibnamefont {Wang}}, \bibinfo {author} {\bibfnamefont {W.}~\bibnamefont
  {Liu}}, \bibinfo {author} {\bibfnamefont {P.}~\bibnamefont {Li}}, \bibinfo
  {author} {\bibfnamefont {Y.}~\bibnamefont {Fu}}, \bibinfo {author}
  {\bibfnamefont {L.}~\bibnamefont {Xiao}}, \bibinfo {author} {\bibfnamefont
  {J.}~\bibnamefont {Ma}}, \ and\ \bibinfo {author} {\bibfnamefont
  {S.}~\bibnamefont {Jia}},\ }\href {\doibase 10.1364/OE.500605} {\bibfield
  {journal} {\bibinfo  {journal} {Opt. Express}\ }\textbf {\bibinfo {volume}
  {31}},\ \bibinfo {pages} {34470} (\bibinfo {year} {2023})}\BibitemShut
  {NoStop}%
\end{thebibliography}

\end{document}